\documentclass[draftclsnofoot,twoside,onecolumn,letter,12pt]{IEEEtran}
\usepackage{cite}
\usepackage{amsmath,amssymb,amsfonts, mathtools}
\usepackage{graphicx}
\usepackage{textcomp}
\usepackage{xcolor}
\usepackage{comment}
\usepackage{commath}
\usepackage{algorithm,algorithmic}
\usepackage{enumerate}
\usepackage{psfrag,subfigure}
\usepackage{amsbsy,epsfig,bbm,mathrsfs,multirow,amsthm}
\usepackage{float}
\usepackage{mathrsfs}
\usepackage{color}
\usepackage{tabulary}
\usepackage{multirow}
\usepackage{cases}
\usepackage{stfloats}
\usepackage{url}
\usepackage[subfigure]{tocloft}
\usepackage{import}
\usepackage{tikz}

\usepackage{xcolor}

\def\checkmark{\tikz\fill[scale=0.4](0,.35) -- (.25,0) -- (1,.7) -- (.25,.15) -- cycle;}
\DeclareMathOperator*{\argmax}{arg\,max}

\begin{document}

\title{3D UAV Trajectory and Data Collection Optimisation via Deep Reinforcement Learning}

\author{Khoi Khac Nguyen, Trung Q. Duong, Tan Do-Duy, Holger Claussen, and Lajos Hanzo
\thanks{Khoi Khac Nguyen and Trung Q. Duong are with the School of Electronics, Electrical Engineering and Computer Science, Queen’s University Belfast, Belfast BT7 1NN, U.K. (e-mail: \{knguyen02,trung.q.duong\}@qub.ac.uk).}
\thanks{Tan Do-Duy is with Ho Chi Minh City University of Technology and Education, Vietnam (e-mail: tandd@hcmute.edu.vn).}
\thanks{Holger Claussen is with Tyndall National Institute, Dublin, Ireland (e-mail: holger.claussen@tyndall.ie).}
\thanks{Lajos Hanzo is with the School of Electronics and Computer Science, University of Southampton, Southampton, SO17 1BJ, U.K. (e-mail: lh@ecs.soton.ac.uk).}
}

\maketitle

\begin{abstract}
Unmanned aerial vehicles (UAVs) are now beginning to be deployed for enhancing the network performance and coverage in wireless communication. However, due to the limitation of their on-board power and flight time, it is challenging to obtain an optimal resource allocation scheme for the UAV-assisted Internet of Things (IoT). In this paper, we design a new UAV-assisted IoT systems relying on the shortest flight path of the UAVs while maximising the amount of data collected from IoT devices. Then, a deep reinforcement learning-based technique is conceived for finding the optimal trajectory and throughput in a specific coverage area. After training, the UAV has the ability to autonomously collect all the data from user nodes at a significant total sum-rate improvement while minimising the associated resources used. Numerical results are provided to highlight how our techniques strike a balance between the throughput attained, trajectory, and the time spent. More explicitly, we characterise the attainable performance in terms of the UAV trajectory, the expected reward and the total sum-rate.
\end{abstract}

{\it Keywords-} UAV-assisted wireless network, trajectory, data collection, and deep reinforcement learning.


\section{Introduction}\label{Sec:Intro}

Given the agility of unmanned aerial vehicles (UAVs), they are capable of supporting compelling applications and are beginning to be deployed more broadly. Recently, the UK and Chile authorities proposed to deliver medical support and other essential supplies by using UAVs to vulnerable people in response to Covid-19 \cite{Drone:UK, Drone:Chile}. In \cite{MG:17:SC}, the authors used UAVs for image collection and high-resolution topography exploration. However, given the several limitations of on-board power level and the ability to adapt to changes in the environment, UAVs may not be fully autonomous and can only operate for short flight-durations, unless remote laser-charging is used \cite{QL:16:VTM}. Moreover, due to some challenging tasks such as topographic surveying, data collection or obstacle avoidance, the existing UAV technologies cannot operate in an optimal manner.\par

Wireless networks supported by UAVs constitute a promising technology for enhancing the network performance \cite{HC:06:ICC}. The applications of UAVs in wireless networks span across diverse research fields, such as wireless sensor networks (WSNs) \cite{JG:18:SAC}, caching \cite{CZ:20:CCN},  heterogeneous cellular networks \cite{HW:20:WC}, massive multiple-input multiple-output (MIMO) \cite{HH:20:VT}, disaster communications \cite{TD:19:GLOBECOM, Trung:19:IWCMC} and device-to-device communications (D2D) \cite{MM:16:WC}. For example, in \cite{Long:EAI}, UAVs were deployed to provide network coverage for people in remote areas and disaster zones. UAVs were also used for collecting data in a WSN \cite{JG:18:SAC}. Nevertheless, the benefits of UAV-aided wireless communication are critically dependent on the limited on-board power level. Thus, the resource allocation of UAV-aided wireless networks plays a pivotal role in approaching the optimal performance. Yet, the existing contributions typically assume having static environment \cite{TD:19:GLOBECOM, Trung:19:IWCMC, Minh:19:WCL} and often ignore the stringent flight time constraints in real-life applications \cite{JG:18:SAC, HW:20:WC, Xiaowei:19:VT}.

Machine learning has recently been proposed for the intelligent support of UAVs and other devices in the network \cite{HZ:20:VT, Xiao:19:VT,KK:19:Access, Khoi:19:Access, Khoi:20:Access, KL:19:VT, UC:19:WC, HH:20:VT, XL:19:VT, CW:19:VT}. Reinforcement learning (RL) is capable of searching for an optimal policy by trial-and-error learning. However, it is challenging for model-free RL algorithms, such as Q-learning to obtain an optimal strategy, while considering a large state and action space. Fortunately, with the emerging neural networks, the sophisticated combination of RL and deep learning, namely deep reinforcement learning (DRL) is eminently suitable for solving high-dimensional problems. Hence, DRL algorithms have been widely applied in fields such as robotics \cite{SG:17:ICRA}, business management \cite{QC:18:AAAI} and gaming \cite{Mnih:13}. Recently, DRL has also become popular in solving diverse problems in wireless networks thanks to their decision-making ability and flexible interaction with the environment \cite{YY:19:SAC, KK:19:Access, Khoi:19:Access, Khoi:20:Access,  KL:19:VT, UC:19:WC, HH:20:VT, XL:19:VT, CW:19:VT, CZ:20:CCN,  NZ:19:WC,  SY:19:VT}. For example, DRL was used for solving problems in the areas of resource allocation \cite{NZ:19:WC, KK:19:Access, Khoi:19:Access}, navigation \cite{DY:18:VT, HH:20:VT} and interference management \cite{UC:19:WC}.

\subsection{Related Contributions}\label{Sec:Works}

UAV-aided wireless networks have also been used for machine-to-machine communications \cite{HW:19:WC} and D2D scenarios in 5G \cite{Minh:19:WCL, Huy:20:CCN}, but the associated resource allocation problems remain challenging in real-life applications. Several techniques  have been developed for solving resource allocation problems \cite{LL:19:WC, LX:19:IOT, KK:19:Access, Khoi:19:Access, DY:18:VT, LN:19:SPAWC}.  In \cite{LL:19:WC}, the authors have conceived a multi-beam UAV communications and a cooperative interference cancellation scheme for maximising the uplink sum-rate received from multiple UAVs by the base stations (BS) on the ground. The UAVs were deployed as access points to serve several ground users in \cite{LX:19:IOT}. Then, the authors proposed successive convex programming for maximising the minimum uplink rate gleaned from all the ground users. In \cite{DY:18:VT}, the authors characterised the tradeoffs between the ground terminal transmission power and the specific UAV trajectory both in a straight and in a circular trajectory.

The issues of data collection, energy minimisation, and path planning have been considered in \cite{QW:18:WC,CZ:18:WCL, HW:18:WCL, HW:19:WC, XL:19:VT, ZW:20:ITJ,JL:20:ITJ,MS:20:WC,  MH:20:C,CZ:20:C}. In \cite{CZ:18:WCL}, the authors minimised the energy consumption of the data collection task considered by jointly optimising the sensor nodes' wakeup schedule and the UAV trajectory. The authors of \cite{HW:18:WCL} proposed an efficient algorithm for joint trajectory and power allocation optimisation in UAV-assisted networks to maximise the sum-rate during a specific length of time. A pair of near-optimal approaches for optimal trajectory was proposed for a given UAV power allocation and power allocation optimisation for a given trajectory. In \cite{HW:19:WC}, the authors introduced a communication framework for UAV-to-UAV communication under the constraints of the UAV's flight speed, location uncertainty and communication throughput. Then, a path planning algorithm was proposed for minimising the associated completion time task while balancing the performance by computational complexity trade-off. However, these techniques mostly operate in offline modes and may impose excessive delay on the system. It is crucial to improve the decision-making time for meeting the stringent requirements of UAV-assisted wireless networks.

Again, machine learning has been recognised as a powerful tool of solving the high-dynamic trajectory and resource allocation problems in wireless networks. In \cite{LN:19:SPAWC}, the authors proposed a model based on the classic k-means algorithm for grouping the users into clusters and assigned a dedicated UAV to serve each cluster. By relying on their decision-making ability, DRL algorithms have been used for lending each node some degree of autonomy \cite{YY:19:SAC, KK:19:Access, Khoi:19:Access, Khoi:20:Access, KL:19:VT, CZ:20:CCN, NZ:19:WC}. In \cite{YY:19:SAC}, an optimal DRL-based channel access strategy to maximise the sum rate and $\alpha$-fairness was considered. In \cite{KK:19:Access, Khoi:19:Access}, we deployed DRL techniques for enhancing the energy-efficiency of D2D communications. In \cite{KL:19:VT}, the authors characterised the DQL algorithm for minimising the data packet loss of UAV-assisted power transfer and data collection systems. As a further advance, caching problems were considered in \cite{CZ:20:CCN} to maximise the cache success hit rate and to minimise the transmission delay. The authors designed both a centralised and a decentralised system model and used an actor-critic algorithm to find the optimal policy.

\begin{table}[h!]
	\centering 
	\caption{A comparison with existing literature}
	\centering 
	\label{Tab:compare}
	\centering
	\begin{tabular}{|c|c|c|c|c|c|c|c|c|c|c|c|c|c|c|}
		\hline
		 & \cite{QW:18:WC} &\cite{CZ:18:WCL}& \cite{JG:18:SAC}& \cite{KL:19:VT}&\cite{XL:19:VT}& \cite{ZW:20:ITJ} & \cite{HH:20:VT} &\cite{JL:20:ITJ} &\cite{MS:20:VT}&\cite{MS:20:WC} &\cite{MH:20:C} &\cite{CZ:20:C} & Our work \\
		\hline
		Trajectory design & \checkmark &  \checkmark &   && \checkmark &\checkmark& \checkmark&  \checkmark&\checkmark&\checkmark &\checkmark&\checkmark &\checkmark\\ \hline
		3D trajectory &  &   &   &  & \checkmark & & &\checkmark&&&\checkmark&&\checkmark\\ \hline
		Uplink &  &\checkmark   & \checkmark  &\checkmark  &  &\checkmark & &&\checkmark&\checkmark&\checkmark&\checkmark&\checkmark\\ \hline
		Downlink & \checkmark &   &   &  &\checkmark  & & &&&&\checkmark&&\\ \hline
		Sum-rate maximisation & \checkmark & & &\checkmark&\checkmark&&&&\checkmark&\checkmark&\checkmark&&\checkmark\\ \hline
		Energy optimisation &  &  \checkmark &   &  &  & \checkmark& &&&&\checkmark&\checkmark&\\ \hline
		Time minimisation & & &\checkmark  & \checkmark&&&&\checkmark&&\checkmark&&&\checkmark\\ \hline
		Dynamic environment & &  &  & &\checkmark&\checkmark&&&\checkmark&&&&\checkmark\\ \hline
		Simple environment & \checkmark&  \checkmark& \checkmark & &&\checkmark&&&&\checkmark&\checkmark&\checkmark&\checkmark\\ \hline
		Complex environment &\checkmark &  &\checkmark  & \checkmark&\checkmark&&\checkmark&\checkmark&\checkmark&\checkmark&\checkmark&&\checkmark\\ \hline
		Mathematical solution  & \checkmark&\checkmark  &\checkmark  & &&\checkmark&&\checkmark&&\checkmark&\checkmark&\checkmark&\\ \hline
		Reinforcement learning & & &  &\checkmark &\checkmark&&\checkmark&&\checkmark&&&&\checkmark\\ \hline
		Deep neural networks & &  & &\checkmark &&&\checkmark&&\checkmark&&&&\checkmark\\ \hline
	\end{tabular}
\end{table}

DRL algorithms have also been applied for path planning in UAV-assisted wireless communications \cite{UC:19:WC, SY:19:VT, HH:20:VT, XL:19:VT, CW:19:VT,  MS:20:VT}. In \cite{UC:19:WC}, the authors proposed a DRL algorithm based on the echo state network of \cite{HJ:01:GMD} for finding the flight path, transmission power and associated cell in UAV-powered wireless networks. The so-called deterministic policy gradient algorithm of \cite{Lillicrap:15} was invoked for UAV-assisted cellular networks in \cite{SY:19:VT}. The UAV's trajectory was designed for maximising the uplink sum-rate attained without the knowledge of the user location and the transmit power. Moreover, in \cite{HH:20:VT}, the authors used the DQL algorithm for the UAV's navigation based on the received signal strengths estimated by a massive MIMO scheme. In \cite{XL:19:VT}, Q-learning was used for controlling the movement of multiple UAVs in a pair of scenarios, namely for static user locations and for dynamic user locations under a random walk model. However, the aforementioned contributions have not addressed the joint trajectory and data collection optimisation of UAV-assisted networks, which is a difficult research challenge. Furthermore, these existing works mostly neglected interference, 3D trajectory and dynamic environment.

\subsection{Contributions and Organisation}
In this paper, we consider a system model relying on a single UAV to serve several user nodes. The UAV is considered to be an information-collecting robot aiming for collecting the maximum amount of data from the users with the shortest distance travelled. We conceive a solution based on the DRL algorithm to find the optimal path of a UAV for maximising the joint reward function based on the shortest flight distance and the uplink transmission rate. We compare the difference between our proposed approach and other existing works in Table \ref{Tab:compare}. Our main contributions are summarised as follows:
\begin{itemize}
	\item The UAV system considered has stringent constraints owing to the position of the destination, the UAV's limited flight time and the communication link's constraint. The UAV's objective is to find an optimal trajectory for maximising the total network throughput, while minimising its distance travelled.
	\item We propose DRL techniques for solving the above problem. The area is divided into a grid to enable fast convergence. Following its training, the UAV can have the autonomy to make a decision concerning its next action at each position in the area, hence eliminating the need for human navigation. This makes UAV-aided wireless communications more reliable, practical and optimises the resource consumption.
	\item Two scenarios are considered relying either on three or five clusters for qualifying the efficiency of our approach in terms of both the sum-rate, the trajectory and the associated time.
\end{itemize}

The rest of our paper is organised as follows. In Section \ref{Sec:Model}, we describe our data collection system model and the problem formulation of IoT networks relying on UAVs. Then, the mathematical background of the DRL algorithms is presented in Section \ref{Sec:BG}. Deep Q-learning (DQL) is employed for finding the best trajectory and for solving our data collection problem in Section \ref{Sec:Alg1}. Furthermore, we use the dueling DQL algorithm of \cite{Wang:15} for improving the system performance and convergence speed in Section \ref{Sec:Alg2}. Next, we characterise the efficiency of the DRL techniques in Section \ref{Sec:Results}. Finally, in Section \ref{Sec:Con}, we summarise our findings and discuss our future research.

\section{System Model and Problem Formulation}\label{Sec:Model}

\begin{figure}[h!]
	\centering
	\subfigure{\includegraphics[width=0.5\textwidth]{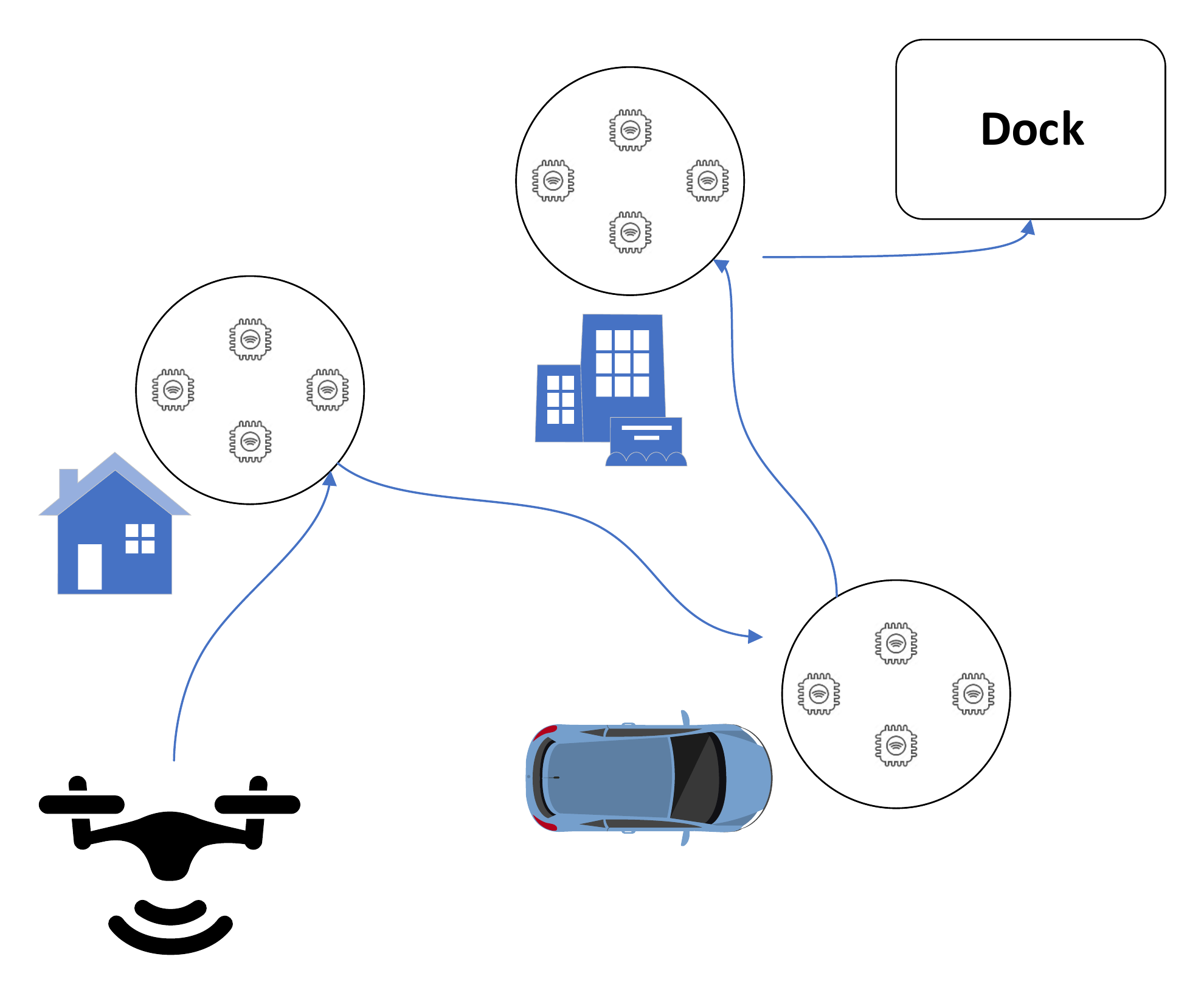}}
	\caption{System model of UAV-aided IoT communications.}
	\label{fig:System}
\end{figure}

Consider a system consisting of a single UAV and $M$ groups of users, as shown in Fig.~\ref{fig:System}, where the UAV relying on a single antenna visits all clusters to cover all the users. The 3D coordinate of the UAV at time step $t$ is defined as $X^t = (x_0^t, y_0^t, H_0^t)$. Each cluster consists of $K$ users, which are unknown and distributed randomly within the coverage radius of $C$. The users are moving following the random walk model with the maximum velocity $v$. The position of the $k$th user in the $m$th cluster at time step $t$ is defined as $X_{m,k}^t = (x_{m,k}^t, y_{m,k}^t)$. The UAV's objective is to find the best trajectory while covering all the users and to reach the dock upon completing its mission.

\subsection{Observation model}

The distance from the UAV to user $k$ in cluster $m$ at time step $t$ is given by:
\begin{equation}
d_{m, k}^t = \sqrt{(x_0^t - x_{m, k}^t) ^ 2 + (y_0^t - y_{m, k}^t) ^2 + {H_0^t} ^2}.
\end{equation}

We assume that the communication channels between the UAV and users are dominated by line-of-sight (LoS) links; thus the channel between the UAV and the $k$th user in the $m$th cluster at time step $t$ follows the free-space path loss model, which is represented as
\begin{equation}
\begin{split}
h_{m, k}^t &= \beta_0 {d_{m, k}^t} ^ {-2} \\
             &= \frac{\beta_0}{(x_0^t - x_{m, k}^t) ^ 2 + (y_0 - y_{m, k}^t) ^2 + {H_0^t}^2},
\end{split}
\end{equation}
where the channel's power gain at a reference distance of $d=1m$ is denoted by $\beta_0$.

The achievable throughput from the $k$th user in the $m$th cluster to the UAV at time $t$ if the user satisfies the distance constraint is defined as follows:
\begin{equation}
R_{m,k}^t = B \log_2 \Bigg(1+ \frac{p_{m, k}^t h_{m, k}^t}{\sum_{i \neq m}^M \sum_j^K p_{i, j}^t h_{i, j}^t + \sum_{u \neq k}^K p_{m, u}^t h_{m, u}^t +  \alpha^2}\Bigg), \forall m, k,
\end{equation}
where $B$ and $\alpha^2$ are the bandwidth and the noise power, respectively. Then the total sum-rate over the $T$ time step from the $k$th user in cluster $m$ to the UAV is given by:
\begin{equation}
R_{m,k} = \int_0^T R_{m,k}^t dt, \forall m,k.
\end{equation}

\subsection{Game formulation}
Both the current location and the action taken jointly influence the rewards obtained by the UAV; thus the trial-and-error based learning task of the UAV satisfies the Markov property. We formulate the associated Markov decision process  (MDP) \cite{Puterman:14} as a 4 tuple $< \mathcal{S}, \mathcal{A}, \mathcal{P}_{ss'}, \mathcal{R} >$, where $\mathcal{S}$ is the state space of the UAV, $\mathcal{A}$ is the action space; $\mathcal{R}$ is the expected reward of the UAV and $\mathcal{P}_{ss'}$ is the probability of transition from state $s$ to state $s'$, where we have $s' = s^{t+1}| s = s^t$. Through learning, the UAV can find the optimal policy $\pi^* : \mathcal{S} \rightarrow \mathcal{A}$ for maximising the reward $\mathcal{R}$. More particularly, we formulate the trajectory and data collection game of UAV-aided IoT networks as follows:
\begin{itemize}
	\item \textit{Agent}: The UAV acts like an agent interacting with the environment to find the peak of the reward.
	
	\item \textit{State space}: We define the state space by the position of UAV as
	\begin{equation}
	\mathcal{S} = \{x, y, H\}.
	\end{equation}
	At time step $t$, the state of the UAV is defined as $s^t = ( x^t, y^t, H^t )$.
	
	\item \textit{Action space}: The UAV at state $s^t$ can choose an action $a^t$ of the action space by following the policy at time-step $t$. By dividing the area into a grid, we can define the action space as follows:
	\begin{equation}
	\mathcal{A} = \{ \text{left}, \text{right}, \text{forward}, \text{backward}, \text{upward}, \text{downward}, \text{hover} \}.
	\end{equation}
The UAV moves in the environment and begins collecting information when the users are in the coverage of the UAV. When the UAV has sufficient information $R_{m,k} \ge r_{min}$ from the $k$th user in the $m$th cluster, that user will be marked as collected in this mission and may not be visited by the UAV again.
	\item \textit{Reward function}: In joint trajectory and data collection optimisation, we design the reward function to be dependent on both the total sum-rate of ground users associated with the UAV plus the reward gleaned when the UAV completes one route, which is formulated as follows:
	\begin{equation}\label{equ:reward}
	R = \frac{\beta}{MK}\left(\sum_m^M \sum_k^K P(m,k)  R_{m, k}\right) + \zeta R_{plus},
	\end{equation}
	where $\beta$ and $\zeta$ are positive variables that represent the trade-off between the network's sum-rate and UAV's movement, which will be described in the sequel. Here, $P(m, k) = \{0,1\}$ indicates whether or not user $k$ of cluster $m$ is associated with the UAV; $R_{plus}$ is the acquired reward when the UAV completes a mission by reaching the final destination. On the other hand, the term $ \frac{\sum_m^M \sum_k^K P(m,k) R_{m, k}}{MK}$ defines the average throughput of all users.
	\item \textit{Probability}: We define $\mathcal{P}_{s^t s^{t+1}}(a^t, \pi)$ as the probability of transition from state $s^t$ to state $s^{t+1}$ by taking the action $a^t$ under the policy $\pi$.
\end{itemize}

At each time step $t$, the UAV chooses the action $a^t$ based on its local information to obtain the reward $r^t$ under the policy $\pi$. Then the UAV moves to the next state $s^{t+1}$ by taking the action $a^t$ and starts collecting information from the users if any available node in the network satisfies the distance constraint. Again, we use the DRL techniques to find the optimal policy $\pi^*$ for the UAV to maximise the reward attained (\ref{equ:reward}). Following the policy $\pi$, the UAV forms a chain of actions $( a^0, a^1, \dots, a^t, \dots, a^{final} )$ to reach the landing dock.

Our target is to maximise the reward expected by the UAV upon completing a single mission during which the UAV flies from the initial position over the clusters and lands at the destination. Thus, we design the trajectory reward $R_{plus}$ when the UAV reaches the destination in two different ways. Firstly, the binary reward function is defined as follows:
\begin{equation}\label{equ:Rplus1}
R_{plus} = \left\{ \begin{array}{rcl} 1 & \mbox{,} & X_{final} \in X_{target} \\
0 & \mbox{,} &  \mbox{otherwise.} \end{array} \right. ,
\end{equation}
where $X_{final}$ and $X_{target}$ are the final position of UAV and the destination, respectively. However, the UAV has to move a long distance to reach the final destination. It may also be trapped in a zone and cannot complete the mission. These situations lead to increased energy consumption and reduced convergence. Thus, we consider the value of $R_{plus}^t$ in a different form by calculating the horizontal distance between the UAV and the final destination at time step $t$, yielding:
\begin{equation}\label{equ:Rplus2}
R^t_{plus} = \left\{ \begin{array}{rcl} 1 & \mbox{,} & X_{final} \in X_{target} \\
\Big (\exp \sqrt{(x_{target}-x_{0}^t)^2 + (y_{target} - y_{0}^t)^2} \Big)^{-1} & \mbox{,} &  \mbox{otherwise.} \end{array} \right.
\end{equation}

When we design the reward function as in (\ref{equ:Rplus2}), the UAV is motivated to move ahead to reach the final destination. However, one of the disadvantages is that the UAV only moves forward. Thus, the UAV is unable to attain the best performance in terms of its total sum-rate in some environmental settings. We compare the performance of the two trajectory reward function definitions in Section \ref{Sec:Results} to evaluate the pros and cons of each approach.

We design the reward function by arranging for a trade-off game with parameters $\beta, \zeta$ to make our approach more adaptive and flexible. By modifying the value of $\beta/\zeta$ , the UAV adapts to several scenarios: a) fast deployment for emergency services, b) maximising the total sum-rate, and c) maximising the number of connections between the UAV and users. Depending on the specific problems, we can adjust the value of the trade-off parameters $\beta, \zeta$ to achieve the best performance. Thus, the game formulation is defined as follows:
\begin{equation}
\begin{split}
\max R =  \quad & \frac{\beta}{MK} \left(\sum_m^M \sum_k^K P(m,k) R_{m, k}\right) + \zeta R_{plus},\\
s.t. \quad&
X_{final} = X_{target},\\
&d_{m, k} \le d_{cons},\\
&R_{m,k} \ge r_{min},\\
&P(m, k) = \{0, 1\},\\
&T \le T_{cons}\\
&\beta \ge 0, \zeta \ge 0,
\end{split}
\end{equation}
where $T$ and $T_{cons}$ are the number of steps that the UAV takes in a single mission and the maximum number of UAV's steps given its limited power, respectively. The distance constraint $d_{m, k} \le d_{cons}$ indicates that the served $(m,k)$-user has a satisfied distance to the UAV. Those stringent constraints, such as the transmission distance, position and flight time make the optimisation problem more challenging. Thus, we propose DRL techniques for the UAV in order to attain the optimal performance.

\section{Preliminaries}\label{Sec:BG}

In this section, we introduce the fundamental concept of Q-learning, where the so-called value function is defined by a reward of the UAV at state $s^t$ as follows:
\begin{equation}
V(s, \pi) = \mathbb{E} \bigg[ \sum_t^T \gamma \mathcal{R}^t (s^t, \pi)| s_0 = s\bigg],
\end{equation}
where $\mathbb{E[\centerdot]}$ represents an average of the number of samples and $0 \le \gamma \le 1$ denotes the discount factor. The value function can be rewritten by expoiting the Markov property as follows:
\begin{equation}
V(s, \pi) = \mathbb{E} \bigg[ \mathcal{R}^t(s^t, \pi)\bigg] + \gamma \sum_{s' \in \mathcal{S}} P_{ss'}(a, \pi) V(s', \pi).
\end{equation}

In a finite game, there is always an optimal policy $\pi^*$ that satisfies the Bellman optimality equation \cite{BD:95:Book:v1}
\begin{equation}
\begin{split}
V^* (s, \pi) &= V (s, \pi^*)\\
& = \max_{a \in \mathcal{A}} \Bigg[ {\mathbb{E} \bigg[ \mathcal{R}^t(s^t, \pi^*)\bigg] + \gamma \sum_{s' \in S} P_{ss'}(a, \pi^*) V(s', \pi^*)} \Bigg] .
\end{split}
\end{equation}

The action-value function is obtained, when the agent at state $s^t$ takes action $a^t$ and receives the reward $r^t$ under the agent policy $\pi$. The optimal Q-value can be formulated as:
\begin{equation}\label{equ:Q}
Q^*(s, a, \pi) = {\mathbb{E} \bigg[ \mathcal{R}^t(s^t, \pi^*)\bigg] + \gamma \sum_{s' \in S} P_{ss'}(a, \pi^*) V(s', \pi^*)}.
\end{equation}

The optimal policy $\pi^*$ can be obtained from $Q^*(s, a, \pi)$ as follows:
\begin{equation}\label{equ:V}
V^*(s, \pi) = \max_{a \in \mathcal{A}} Q(s, a, \pi).
\end{equation}

From (\ref{equ:Q}) and (\ref{equ:V}), we have
\begin{equation}
\begin{split}
Q^*(s, a, \pi) \; &  =  \mathbb{E} \bigg[ \mathcal{R}^t(s^t, \pi^*)\bigg] + \gamma \sum_{s' \in S} P_{ss'}(a, \pi^*) \max_{a' \in \mathcal{A}} Q(s', a', \pi), \\
&= \mathbb{E} \bigg[ \mathcal{R}^t(s^t, \pi^*)+ \gamma \max_{a' \in \mathcal{A}} Q(s', a', \pi)\bigg] ,
\end{split}
\end{equation}
where the agent takes the action $a' = a^{t+1}$ at state $s^{t+1}$.

Through learning, the Q-value is updated based on the available information as follows:
\begin{equation}\label{equ:Qupdate}
\begin{split}
Q(s, a, \pi) = \; Q(s, a, \pi) + \alpha \bigg[ \mathcal{R}^t(s^t, \pi^*)
+ \gamma  \max_{a' \in \mathcal{A}} Q(s', a', \pi) - Q(s, a, \pi) \bigg],
\end{split}
\end{equation}
where $\alpha$ denotes the updated parameter of the Q-value function.

In RL algorithms, it is challenging to balance the \textit{exploration} and \textit{exploitation} for appropriately selecting the action. The most common approach relies on the $\epsilon$-greedy policy for the action selection mechanism as follows:
\begin{equation}\label{equ:ep}
a = \left\{ \begin{array}{rcl} \argmax Q(s, a, \pi) & \mbox{with} & \epsilon  \\
\mbox{randomly} & \mbox{if} &  1- \epsilon. \end{array} \right.
\end{equation}

Upon assuming that each episode lasts $T$ steps, the action at time step $t$ is $a^t$ that is selected by following the $\epsilon$-greedy policy as in (\ref{equ:ep}). The UAV at state $s^t$ communicates with the user nodes from the ground if the distance constraint of $d_{m, k} \le d_{cons}$ is satisfied. Following the information transmission phase, the user nodes are marked as collected users and may not be revisited later during that mission. Then, after obtaining the immediate reward $r(s^t, a^t)$ the agent at state $s^t$ takes action $a^t$ to move to state $s^{t+1}$ as well as to update the Q-value function in (\ref{equ:Qupdate}). Each episode ends when the UAV reaches the final destination and the flight duration constraint is satisfied.

\section{An effective deep reinforcement learning approach for  UAV-assisted IoT networks}\label{Sec:Alg1}

In this section, we conceive the DQL algorithm for trajectory and data collection optimisation in UAV-aided IoT networks. However, Q-learning technique typically falters for large state and action spaces due to its excessive Q-table size. Thus, instead of applying the Q-table in Q-learning, we use deep neural networks to represent the relationship between the action and state space. Furthermore, we employ a pair of techniques for stabilising the neural network's performance in our DQL algorithm as follows:
\begin{itemize}
	\item \textit{Experience relay buffer}: Instead of using current experience, we use a so-called relay buffer $\mathcal{B}$ to store the transitions $(s, a, r, s')$ for supporting the neural network in overcoming any potential instability. When the buffer $\mathcal{B}$ is filled with the transitions, we randomly select a mini-batch of $K$ samples for training the networks. The finite buffer size of $\mathcal{B}$ allows it to be always up-to-date, and the neural networks learn from the new samples.
	\item \textit{Target networks}: If we use the same network to calculate the state-action value $Q$ and the target network, the network can be shifted dramatically in the training phase. Thus, we employ a target network $Q'$ for the target value estimator. After a number of iterations, the parameters of the target network $Q'$ will be updated by the network $Q$.
\end{itemize}

\begin{algorithm}[t!]
	\caption{The deep Q-learning algorithm for trajectory and data collection optimisation in UAV-aided IoT networks}
	\begin{algorithmic}[1]
		\label{alg:DQL}
		\STATE Initialise the network $Q$ and the target network $Q'$ with the random parameters $\theta$ and $\theta'$, respectively
		\STATE Initialise the replay memory pool $\mathcal{B}$
		\FOR{episode = $1,\dots, L$}
		\STATE Receive initial observation state $s^0$
		\WHILE{$X_{final} \notin X_{target}$ or $T \le T_{cons}$}
		\STATE Obtain the action $a^t$ of the UAV according to the $\epsilon$-greedy mechanism (\ref{equ:ep})
		\STATE Execuse the action $a^t$ and estimate the reward $r^t$ according to (\ref{equ:reward})
		\STATE Observe the next state $s^{t+1}$
		\STATE Store the transition $(s^t, a^t, r^t, s^{t+1})$ in the replay buffer $\mathcal{B}$
		\STATE Randomly select a mini-batch of $K$ transitions $(s^k, a^k, r^k, s^{k+1})$ from $\mathcal{B}$
		\STATE Update the network parameters using gradient descent to minimise the loss
		\begin{equation}
		\mathbb{L}(\theta) = \mathbb{E}_{s, a, r, s'} \Bigg[\bigg(y^{DQL} - Q(s, a; \theta)\bigg)^2 \Bigg],
		\end{equation}
		The gradient update is
		\begin{equation}
		\nabla_\theta \mathbb{L}(\theta) = \mathbb{E}_{s, a, r, s'} \Bigg[\bigg(y^{DQL} - Q(s, a; \theta)\bigg)\nabla_\theta Q(s, a;\theta) \Bigg],
		\end{equation}
		\STATE Update the state $s^t = s^{t+1}$
		\STATE Update the target network parameters after a number of iterations as $\theta' = \theta$
		\ENDWHILE
		\ENDFOR
	\end{algorithmic}
\end{algorithm}

The neural network parameters are updated by minimising the loss function defined as follows:
\begin{equation}\label{equ:DQLloss}
\mathbb{L}(\theta)= \mathbb{E}_{s, a, r, s'} \Bigg[ \bigg(y^{DQL} - Q(s, a; \theta)\bigg)^2 \Bigg],
\end{equation}
where  $\theta$ is a parameter of the network $Q$ and we have
\begin{equation}
y = \left\{ \begin{array}{rcl} r^t & \mbox{if terminated at} \; s^{t+1}  \\
r^t + \gamma \max_{a' \in \mathcal{A}} Q'(s', a'; \theta') & \mbox{otherwise.} \end{array} \right.
\end{equation}

The details of the DQL approach in our joint trajectory and data collection trade-off game designed for UAV-aided IoT networks are presented in Alg. \ref{alg:DQL} where $L$ denotes the number of episode. Moreover, in this paper, we design the reward obtained in each step to assume one of two different forms and compare them in our simulation results. Firstly, we calculate the difference between the  current and the previous reward of the UAV as follows:
\begin{equation}\label{equ:R1}
r_1^t (s^t, a^t) = r^t (s^t, a^t) - r^{t-1}(s^{t-1}, a^{t-1}).
\end{equation}

Secondly, we design the total episode reward as the accumulation of all immediate rewards of each step within one episode as
\begin{equation}\label{equ:R2}
r_2^t (s^t, a^t) = \sum^t_{i=0} r_1^t(s^t, a^t).
\end{equation}

\section{Deep reinforcement learning approach for  UAV-assisted IoT networks: A dueling deep Q-learning approach }\label{Sec:Alg2}

According to Wang \textit{et. al.} \cite{Wang:15}, the standard Q-learning algorithm often falters due to the over-supervision of all the state-action pairs. On the other hand, it is unnecessary to estimate the value of each action choice in a particular state. For example, in our environment setting, the UAV has to consider moving either to the left or to the right when it hits the boundaries. Thus, we can improve the convergence speed by avoiding visiting all state-action pairs. Instead of using Q-value function of the conventional DQL algorithm, the dueling neural network of \cite{Wang:15} is introduced for improving the convergence rate and stability. The so-called advantage function $A (s, a) = Q(s, a) - V( s) $ related both to the value function and to the Q-value function describes the importance of each action related to each state.

\begin{algorithm}[t!]
	\caption{The dueling deep Q-learning algorithm for trajectory and data collection optimisation in UAV-aided IoT networks}
	\begin{algorithmic}[1]
		\label{alg:DuelingDQL}
		\STATE Initialise the network $Q$ and the target network $Q'$ with the random parameters, $\theta$ and $\theta'$, respectively
		\STATE Initialise the replay memory pool $\mathcal{B}$
		\FOR{episode = $1,\dots, L$}
		\STATE Receive the initial observation state $s^0$
		\WHILE{$X_{final} \notin X_{target}$ or $T \le T_{cons}$}
		\STATE Obtain the action $a^t$ of the UAV according to the $\epsilon$-greedy mechanism (\ref{equ:ep})
		\STATE Execute the action $a^t$ and estimate the reward $r^t$ according to (\ref{equ:reward})
		\STATE Observe the next state $s^{t+1}$
		\STATE Store the transition $(s^t, a^t, r^t, s^{t+1})$ in the replay buffer $\mathcal{B}$
		\STATE Randomly select a mini-batch of $K$ transitions $(s^k, a^k, r^k, s^{k+1})$ from $\mathcal{B}$
		\STATE Estimate the Q-value function by combining the two streams as follows:
		\begin{equation}
		\begin{split}
		Q(s, a; \; \theta, \theta_A, \theta_V) = V(s ;\theta_V) + \Bigg( A(s, a; \theta_A) - \frac{1}{|\mathcal{A}|} \sum_{a'} A(s, a'; \theta_A) \Bigg).
		\end{split}
		\end{equation}
		\STATE Update the network parameters using gradient descent to minimise the loss
		\begin{equation}
		\mathbb{L} (\theta) = \mathbb{E}_{s, a, r, s'} \Bigg[ \bigg(y^{DuelingDQL} - Q(s, a; \theta, \theta_A, \theta_V)\bigg)^2 \Bigg],
		\end{equation}
		\STATE 	where
		\begin{equation}
		y^{DuelingDQL} = r^t +\gamma \max_{a' \in \mathcal{A}} Q'(s', a'; \theta', \theta_A, \theta_V).
		\end{equation}
		\STATE Update the state $s^t = s^{t+1}$
		\STATE Update the target network parameters after a number of iterations as $\theta' = \theta$
		\ENDWHILE
		\ENDFOR
	\end{algorithmic}
\end{algorithm}

The idea of a dueling deep network is based on a combination of two streams of the value function and the advantage function used for estimating the single output $Q$-function. One of the streams of a fully-connected layer estimates the value function $V(s; \theta_V)$, while the other stream outputs a vector $A(s, a; \theta_A)$, where $\theta_A$ and $\theta_V$ represent the parameters of the two networks. The $Q$-function can be obtained by combining the two streams' outputs as follows:
\begin{equation}\label{equ:A}
Q(s, a; \theta, \theta_A, \theta_V) = V( s; \theta_V) + A(s, a; \theta_A).
\end{equation}

Equation (\ref{equ:A}) applies to all $(s, a)$ instances; thus, we have to replicate the scalar $V(s; \theta_V)$, $|\mathcal{A}|$ times to form a matrix. However, $Q(s, a; \theta, \theta_A, \theta_V)$ is a parameterised estimator of the true Q-function; thus, we cannot uniquely recover the value function $V$ and the advantage function $A$. Therefore, (\ref{equ:A}) results in poor practical performances when used directly. To address this problem, we can map the advantage function estimator to have no advantage at the chosen action by combining the two streams as follows:
\begin{equation}\label{equ:Amax}
\begin{split}
Q(s, a; \theta, \theta_A, \theta_V) = V(s ;\theta_V) + \bigg( A(s, a; \theta_A) - \max_{a' \in |\mathcal{A}|} A(s, a'; \theta_A) \bigg).
\end{split}
\end{equation}

Intuitively, for $a^* = \argmax_{a' \in \mathcal{A}}Q(s, a'; \theta, \theta_A, \theta_V) = \argmax_{a' \in \mathcal{A}} A(s, a'; \theta_A)$, we have $\linebreak Q(s, a^*; \theta, \theta_A, \theta_V) = V(s; \theta_V)$. Hence, the stream $V(s; \theta_V)$ estimates the value function and the other streams is the advantage function estimator. We can transform (\ref{equ:Amax}) using an average formulation instead of the \textit{max} operator as follows:
\begin{equation}\label{equ:Aavg}
\begin{split}
Q(s, a; \theta, \theta_A, \theta_V) = V(s ;\theta_V) + \Bigg( A(s, a; \theta_A) - \frac{1}{|\mathcal{A}|} \sum_{a'} A(s, a'; \theta_A) \Bigg).
\end{split}
\end{equation}

Now, we can solve the problem of identifiability by subtracting the mean as in (\ref{equ:Aavg}). Based on (\ref{equ:Aavg}), we propose a dueling DQL algorithm for our joint trajectory and data collection problem in UAV-assisted IoT networks relying on Alg. \ref{alg:DuelingDQL}. Note that estimating $V(s; \theta_V)$ and $A(s, a ; \theta_A)$ does not require any extra supervision and they will be computed automatically.

\section{Simulation Results}\label{Sec:Results}
In this section, we present our simulation results characterising the joint optimisation problem of UAV-assisted IoT networks. To highlight the efficiency of our proposed model and the DRL methods, we consider a pair of scenarios: a simple having three clusters, and a more complex one with five clusters in the coverage area. We use Tensorflow 1.13.1 \cite{Abadi:16} and the Adam optimiser of \cite{DJ:14} for training the neural networks. All the other parameters are provided in Table \ref{tab:Params}.

\begin{table}[t!]
	\renewcommand{\arraystretch}{1.2}
	\caption{SIMULATION PARAMETERS}
	\label{tab:Params}
	\centering
	\begin{tabular}{l|l}
		\hline
		Parameters & Value \\
		\hline
		Bandwidth ($W$)  & $1$ MHz \\
		UAV transmission power & $5$ W \\
		The start position of UAV & $(0, 0, 200)$\\
		Discounting factor & $\gamma = 0.9$\\
		Max number of users per cluster & $10$\\
		Noise power & $\alpha^2 = -110dBm$ \\
		The reference channel power gain & $\beta_0 = -50dB$\\
		Path-loss exponent & $2$ \\
		\hline
	\end{tabular}
\end{table}

\begin{figure}[t!]
	\centering
	\subfigure{\includegraphics[width=0.65\textwidth]{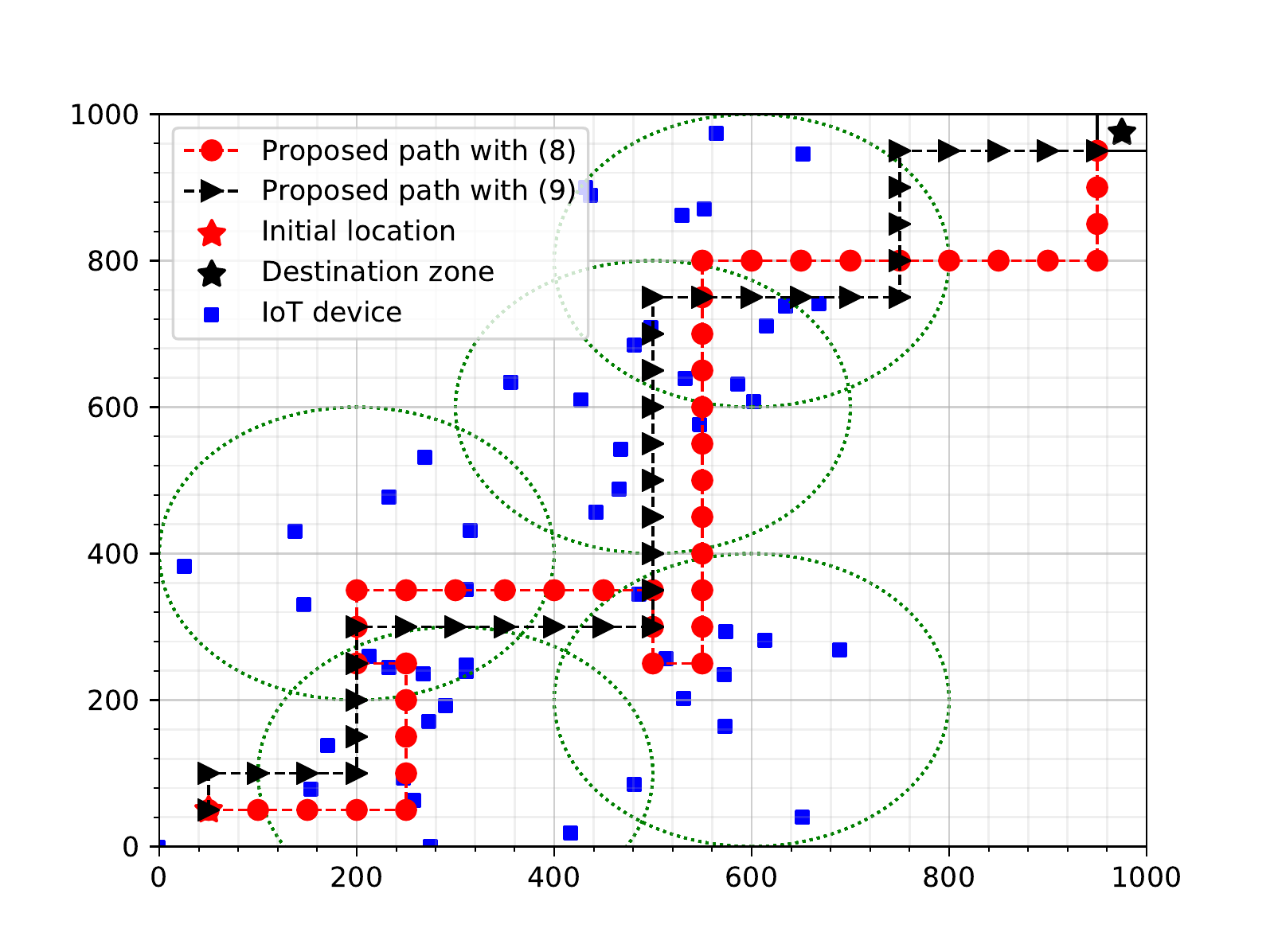}}
	\caption{Trajectory obtained by using our DQL algorithm}
	\label{fig:Traj}
\end{figure}
In Fig. (\ref{fig:Traj}), we present the trajectory obtained after training using the DQL algorithm in the $5$-cluster scenario. The green circle and blue dots represent the clusters' coverage and the user nodes, respectively. The red dots and black triangles in the figure represent the UAV's state after taking action. The UAV starts at $(0, 0)$, visits about $40$ users, and lands at the destination that is denoted by the black square. In a complex environment setting, it is challenging to expect the UAV to visit all users, while satisfying the flight-duration and power level constraints.

\subsection{Expected reward}

\begin{figure}[h!]
	\centering
	\subfigure[]{\includegraphics[width=0.65\textwidth]{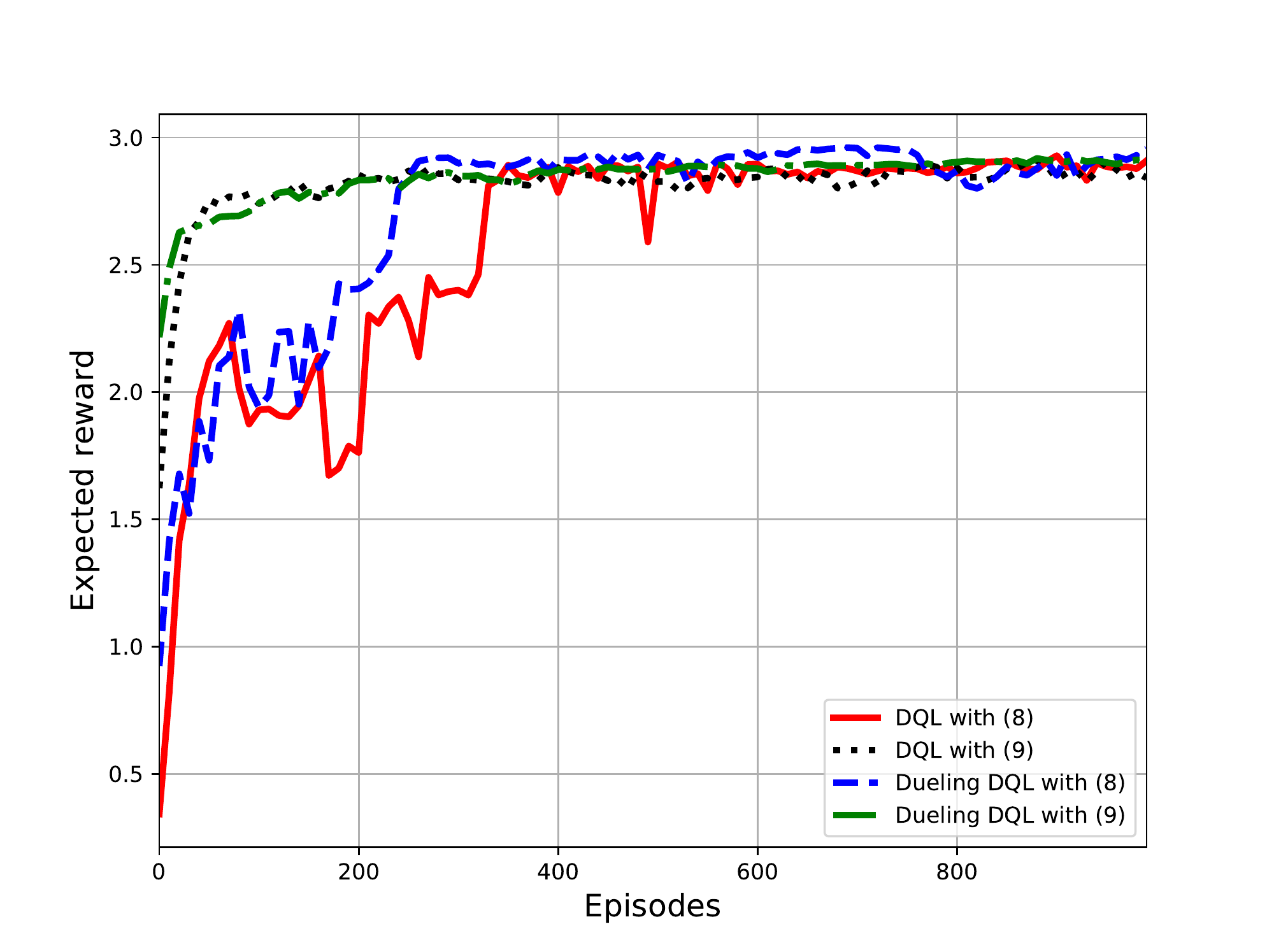}}
	\subfigure[]{\includegraphics[width=0.65\textwidth]{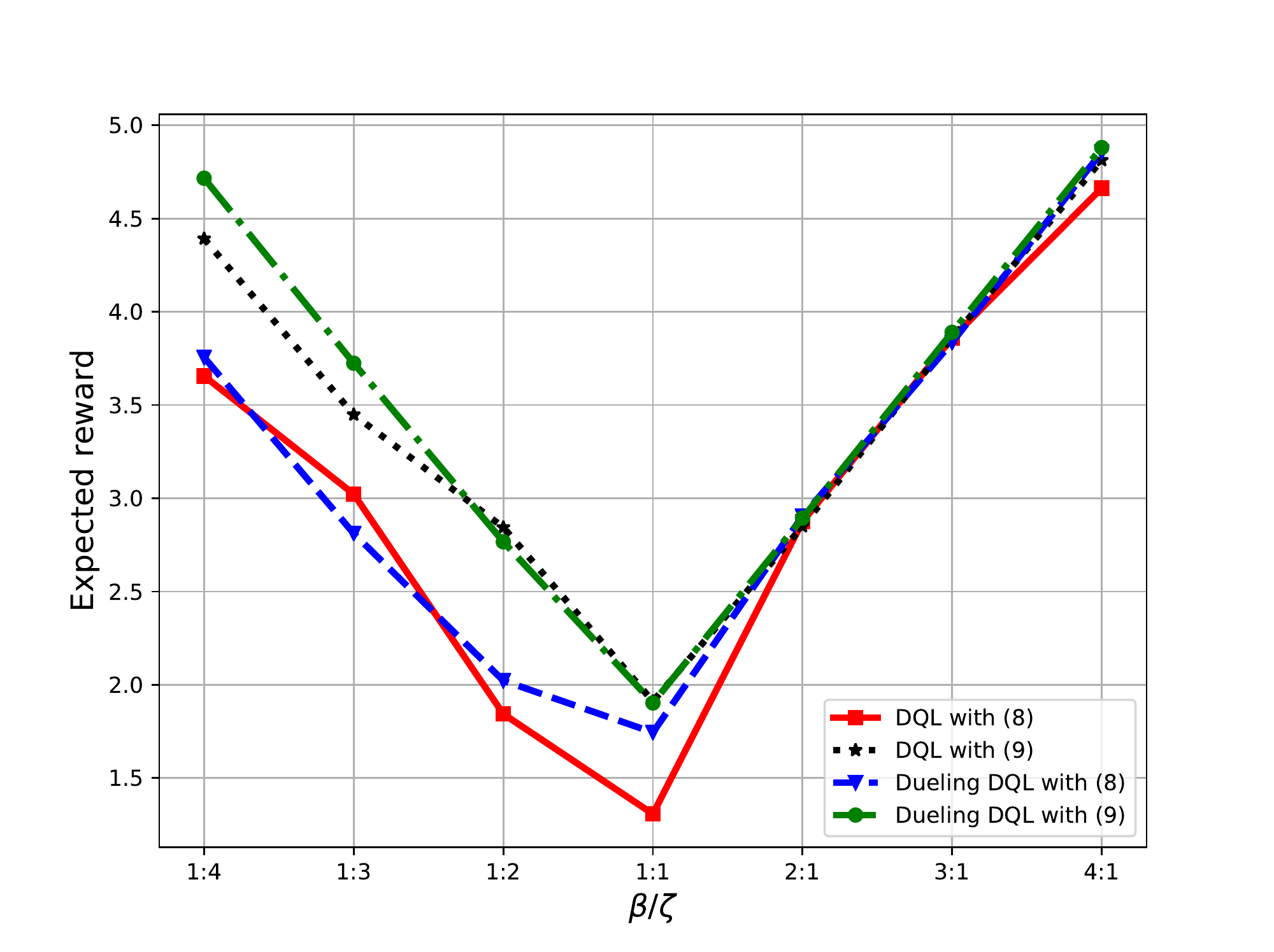}}
	\caption{The performance when using the DQL and dueling DQL algorithms with 3 clusters while considering different $\beta$/$\zeta$ values}
	\label{fig:reward3clusters}
\end{figure}

\begin{figure}[h!]
	\centering
	\subfigure[With (\ref{equ:Rplus1})]{\includegraphics[width=0.65\textwidth]{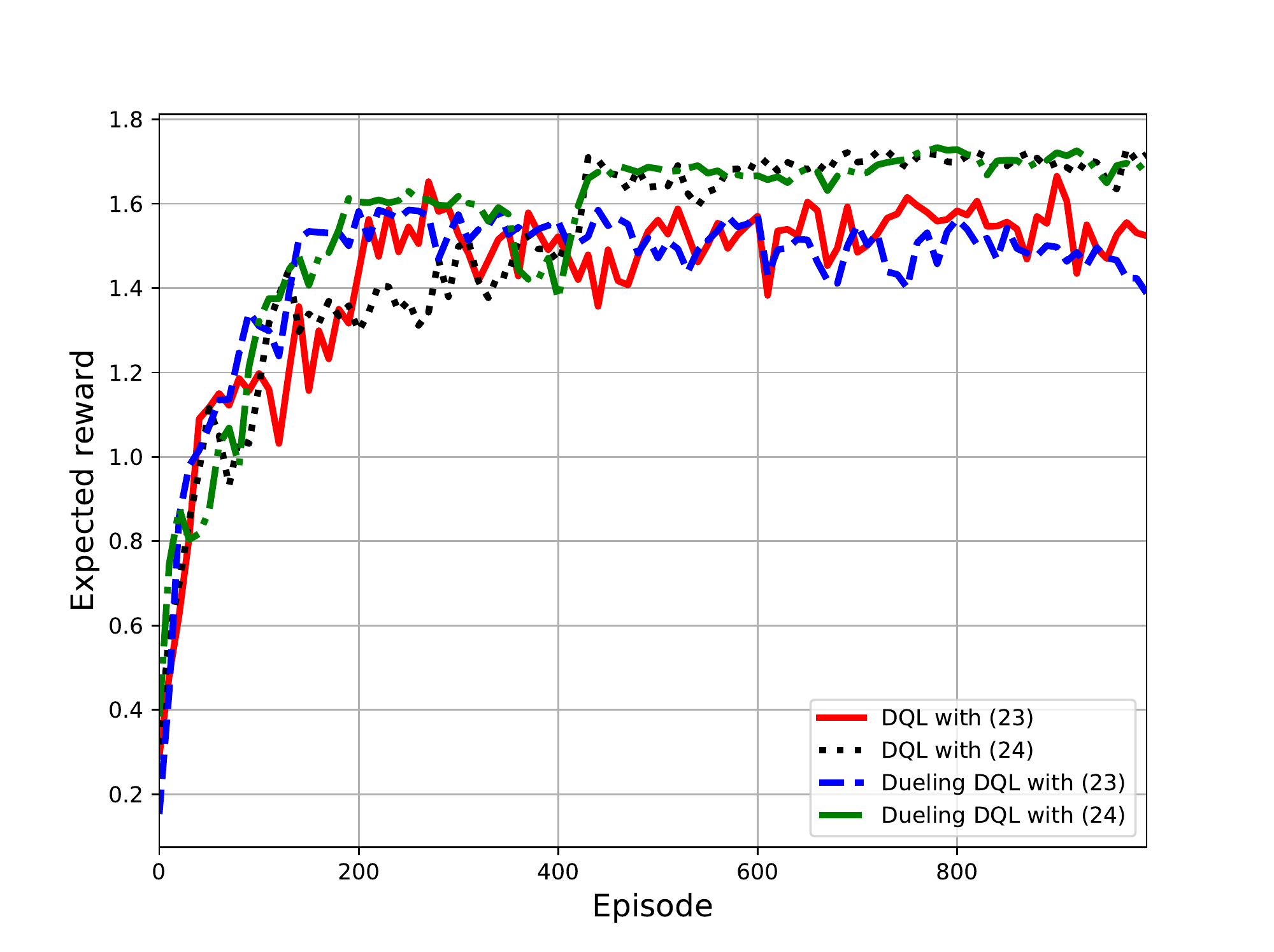}}
	\subfigure[With (\ref{equ:Rplus2})]{\includegraphics[width=0.65\textwidth]{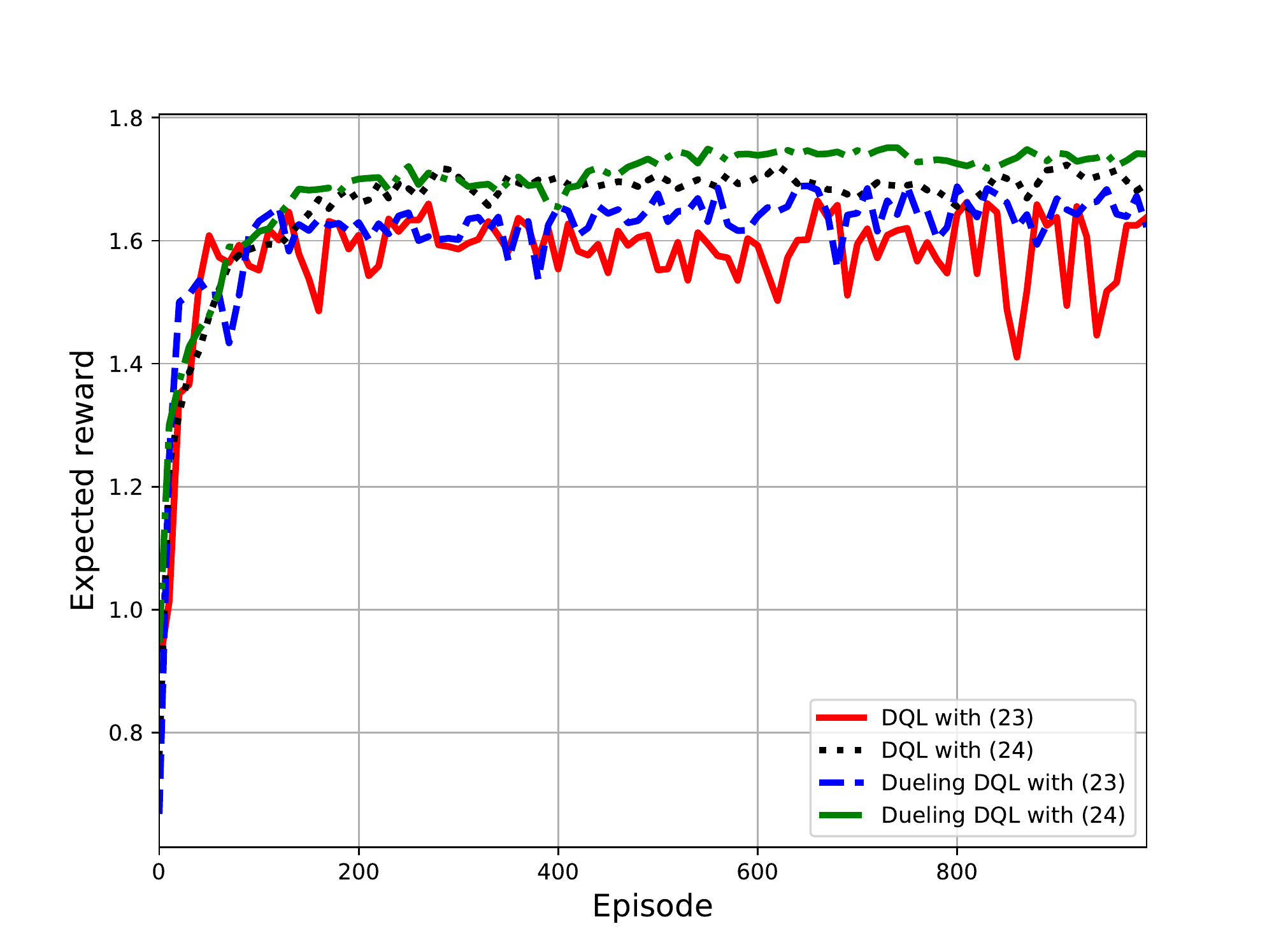}}
	\caption{The expected reward when using the DQL and dueling DQL algorithms with 5-cluster scenario}
	\label{fig:reward5clusters}
\end{figure}

For purposes of comparison, we run the algorithm five times in five different environmental settings and take the average to draw the figures. Firstly, we compare the reward obtained following (\ref{equ:reward}). Let us consider the $3$-cluster scenario and $\beta/\zeta = 2:1$ in Fig. (\ref{fig:reward3clusters}a), where the DQL and dueling DQL algorithms using the exponential function ($\ref{equ:Rplus2}$) reach the best performance. When using the exponential trajectory design function (\ref{equ:Rplus2}), the performance converges faster than that of the DQL and dueling DQL methods using the binary trajectory function (\ref{equ:Rplus1}). In addition, in Fig. (\ref{fig:reward3clusters}b), we compare the performance of the DQL and dueling DQL techniques using different $\beta/\zeta$ values. The average performance of the dueling DQL algorithm is better than that of the DQL algorithm. In conjunction, the results of using the exponential function ($\ref{equ:Rplus2}$) is better than that of the ones using the binary function ($\ref{equ:Rplus1}$).

Furthermore, we compare the rewards obtained by the DQL and dueling DQL algorithms in complex scenarios with $5$ clusters and $50$ user nodes in Fig. (\ref{fig:reward5clusters}). The performance of using the episode reward (\ref{equ:R2}) is better than that using the immediate reward (\ref{equ:R1}) in both trajectory designs relying on the DQL and dueling DQL algorithms. In Fig. (\ref{fig:reward5clusters}a), we compare the performance in conjunction with the binary trajectory design while in Fig. (\ref{fig:reward5clusters}b) the exponential trajectory design is considered. For $\beta/\zeta= 1:1$, the rewards obtained by the DQL and dueling DQL are similar and stable after about $400$ episodes. When using the exponential function (\ref{equ:Rplus2}), the dueling DQL algorithm reaches the best performance. Moreover, the convergence of the dueling DQL technique is faster than that of the DQL algorithm.

\begin{figure}[h!]
	\centering
	\subfigure{\includegraphics[width=0.65\textwidth]{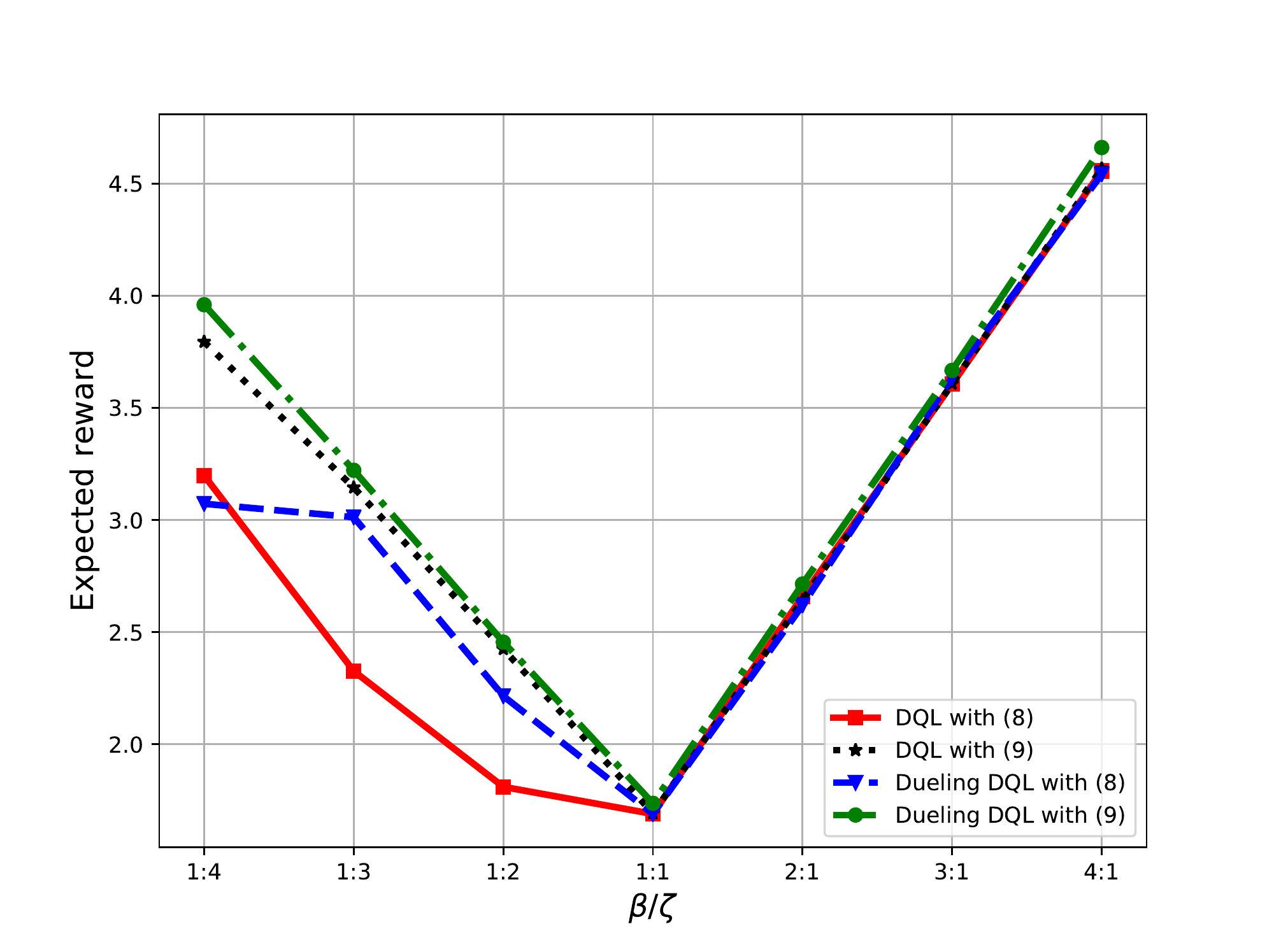}}
	\caption{The performance when using the DQL and dueling DQL algorithms with 5 clusters and different $\beta/\zeta$ values}
	\label{fig:reward5clusters2}
	
\end{figure}

In Fig. (\ref{fig:reward5clusters2}), we compare the performance of the DQL and of the dueling DQL algorithms while considering different $\beta/\zeta$ parameter values. The dueling DQL algorithm shows better performance for all the $\beta/\zeta$ pair values, exhibiting better rewards. In addition, when using the exponential function (\ref{equ:Rplus2}), both proposed algorithms show better performance than the ones using the binary function (\ref{equ:Rplus1}) if $\beta/\zeta \le 1:1$, but it becomes less effective when $\beta/\zeta$ is set higher.

\begin{figure}[h!]
	\centering
	\subfigure[]{\includegraphics[width=0.65\textwidth]{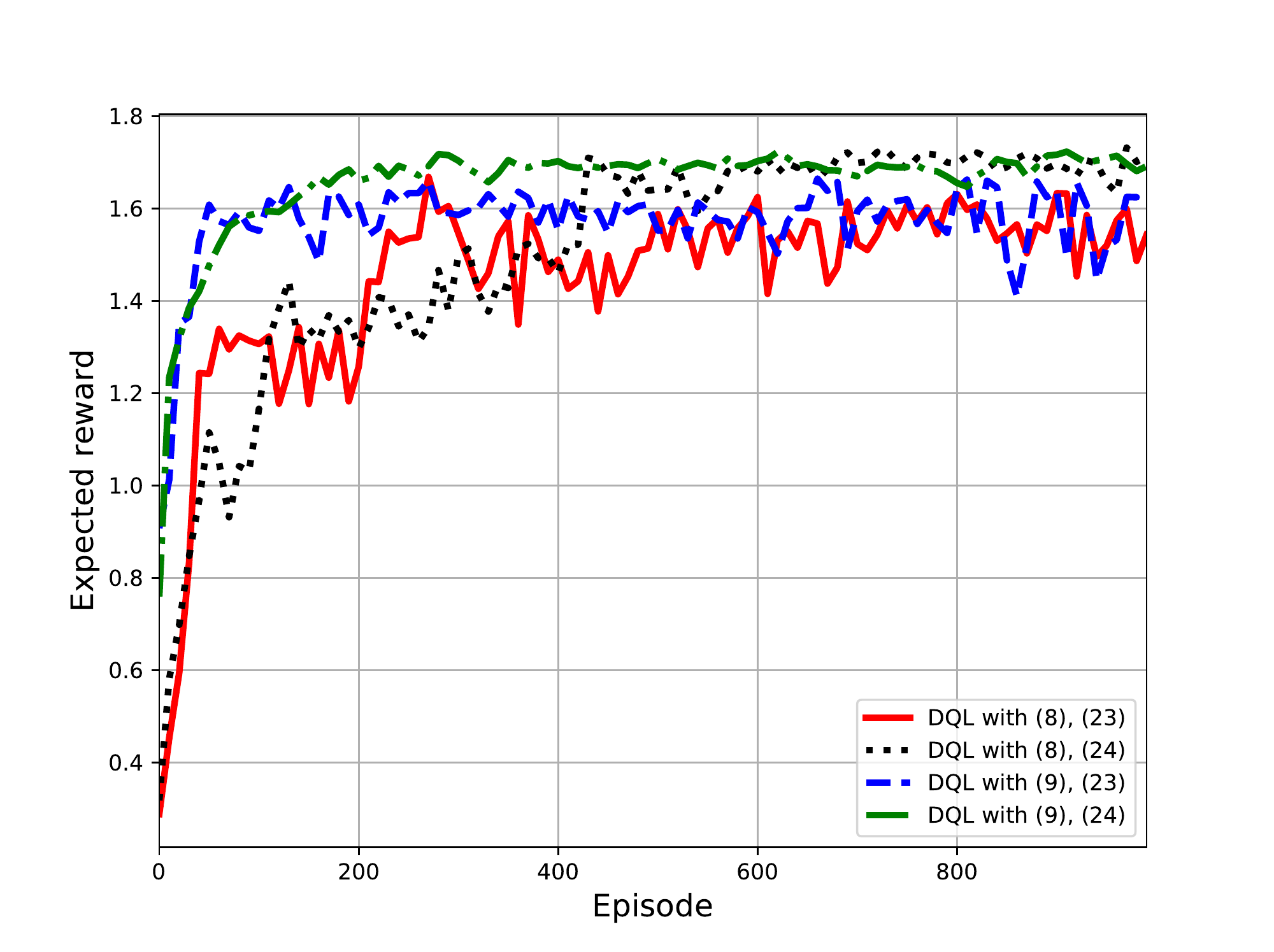}}
	\subfigure[]{\includegraphics[width=0.65\textwidth]{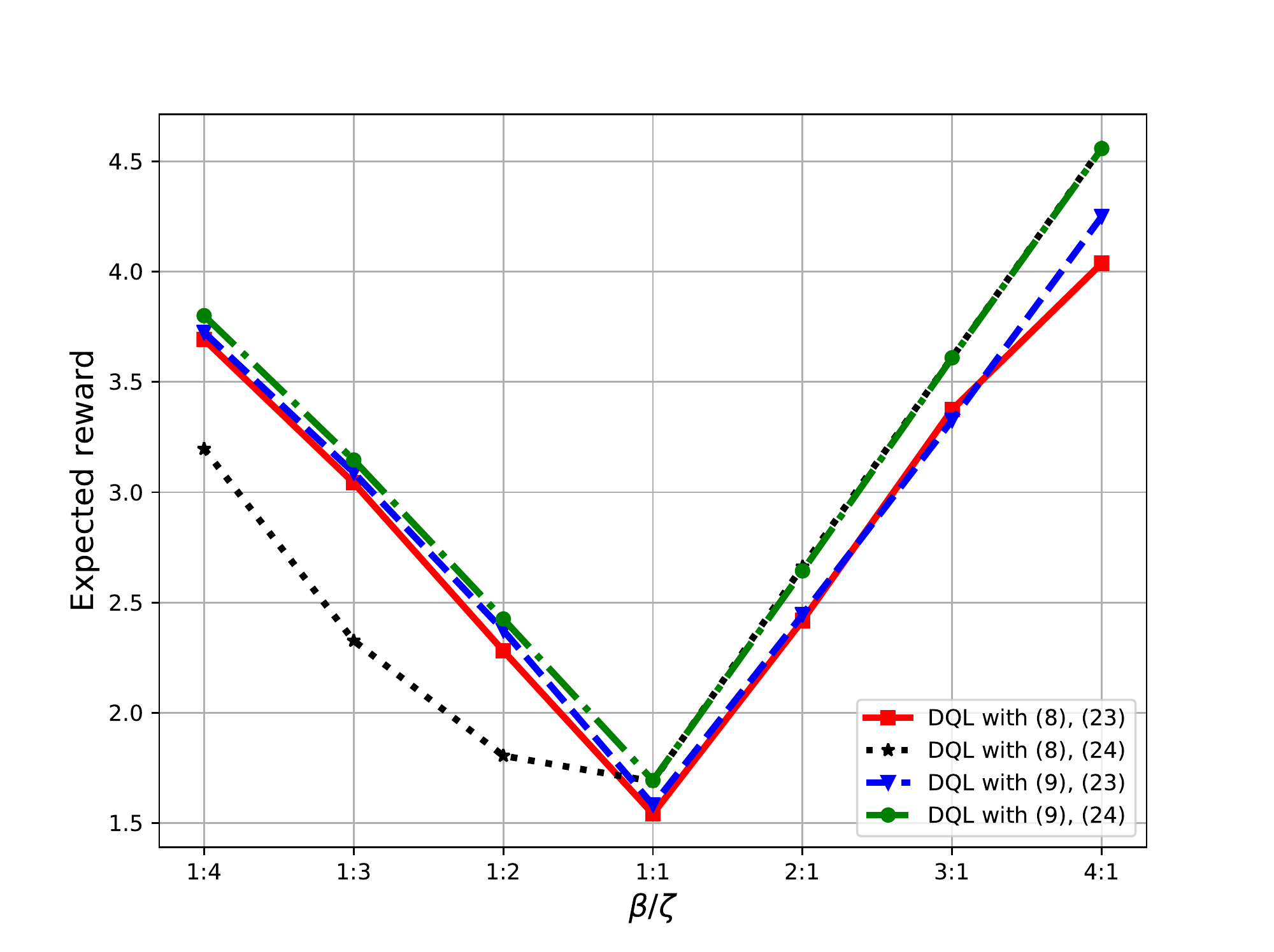}}
	\caption{The expected reward when using the DQL algorithm with 5 clusters and different reward function settings}
	\label{fig:reward5clusters3}
\end{figure}

\begin{figure}[h!]
	\centering
	\subfigure{\includegraphics[width=0.65\textwidth]{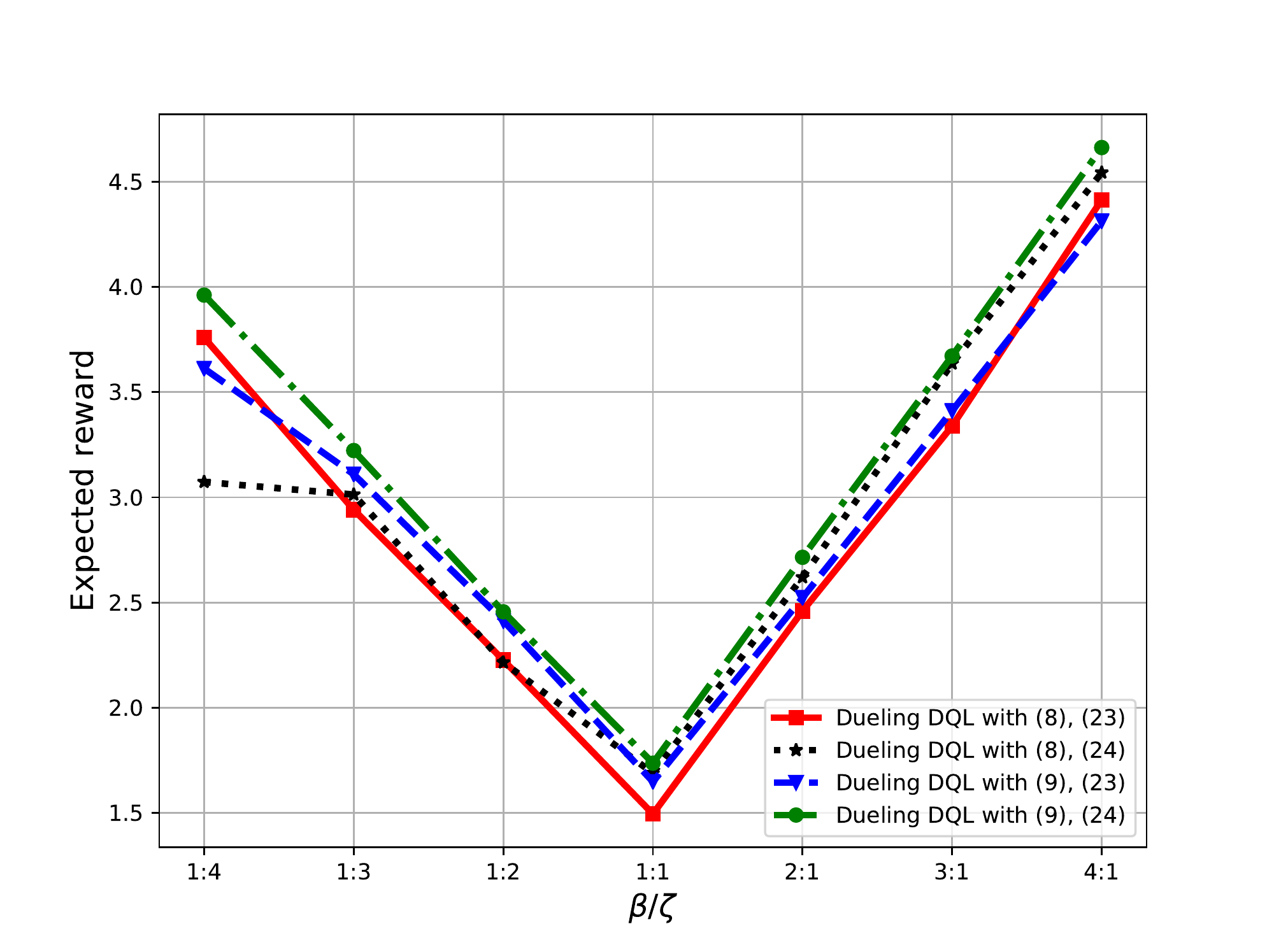}}
	\caption{The performance when using the dueling DQL with 5 clusters,  and different $\beta/\zeta$ values}
	\label{fig:reward5clusters4}
\end{figure}
We compare the performance of the DQL and of the dueling DQL algorithm using different reward function setting in Fig. (\ref{fig:reward5clusters3}) and in Fig. (\ref{fig:reward5clusters4}), respectively. The DQL algorithm reaches the best performance when using the episode reward (\ref{equ:R2}) in Fig. (\ref{fig:reward5clusters3}a) while the fastest convergence speed can be achieved by using the exponential function (\ref{equ:Rplus2}). When $\beta/\zeta \ge 1:1$, the DQL algorithm relying on the episode function (\ref{equ:R2}) outperforms the ones using the immediate reward function (\ref{equ:R1}) in Fig. (\ref{fig:reward5clusters3}b). The reward (\ref{equ:reward}) using the exponential trajectory design (\ref{equ:Rplus2}) has a better performance than that using the binary trajectory design (\ref{equ:Rplus1}) for all the $\beta/\zeta$ values. The similar results are shown when using the dueling DQL algorithm in Fig. (\ref{fig:reward5clusters4}). The immediate reward function (\ref{equ:R1}) is less effective than the episode reward function (\ref{equ:R2}).

\subsection{Throughput comparison}

\begin{figure}[h!]
	\centering
	\subfigure[With (\ref{equ:Rplus1})]{\includegraphics[width=0.65\textwidth]{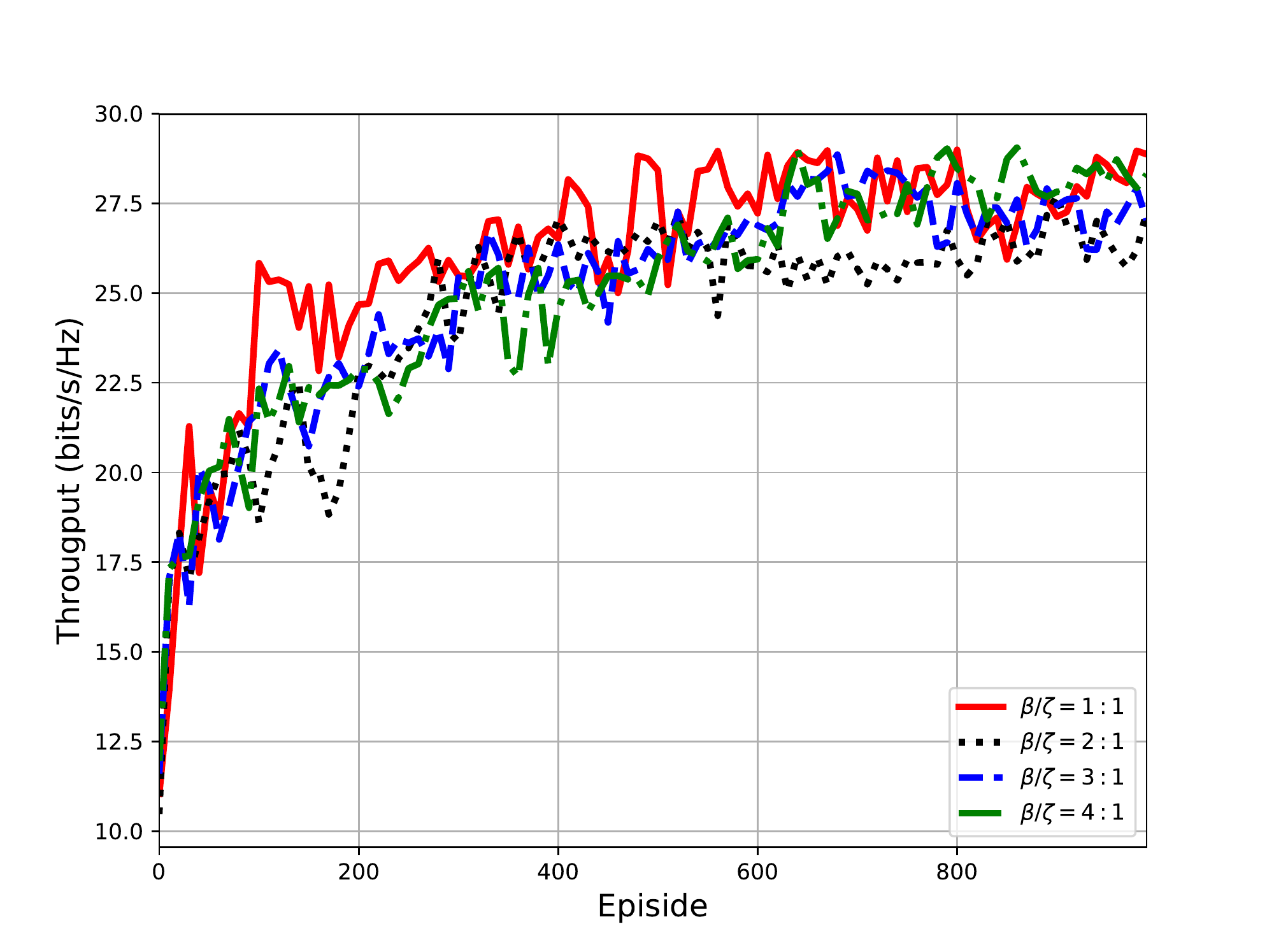}}
	\subfigure[]{\includegraphics[width=0.65\textwidth]{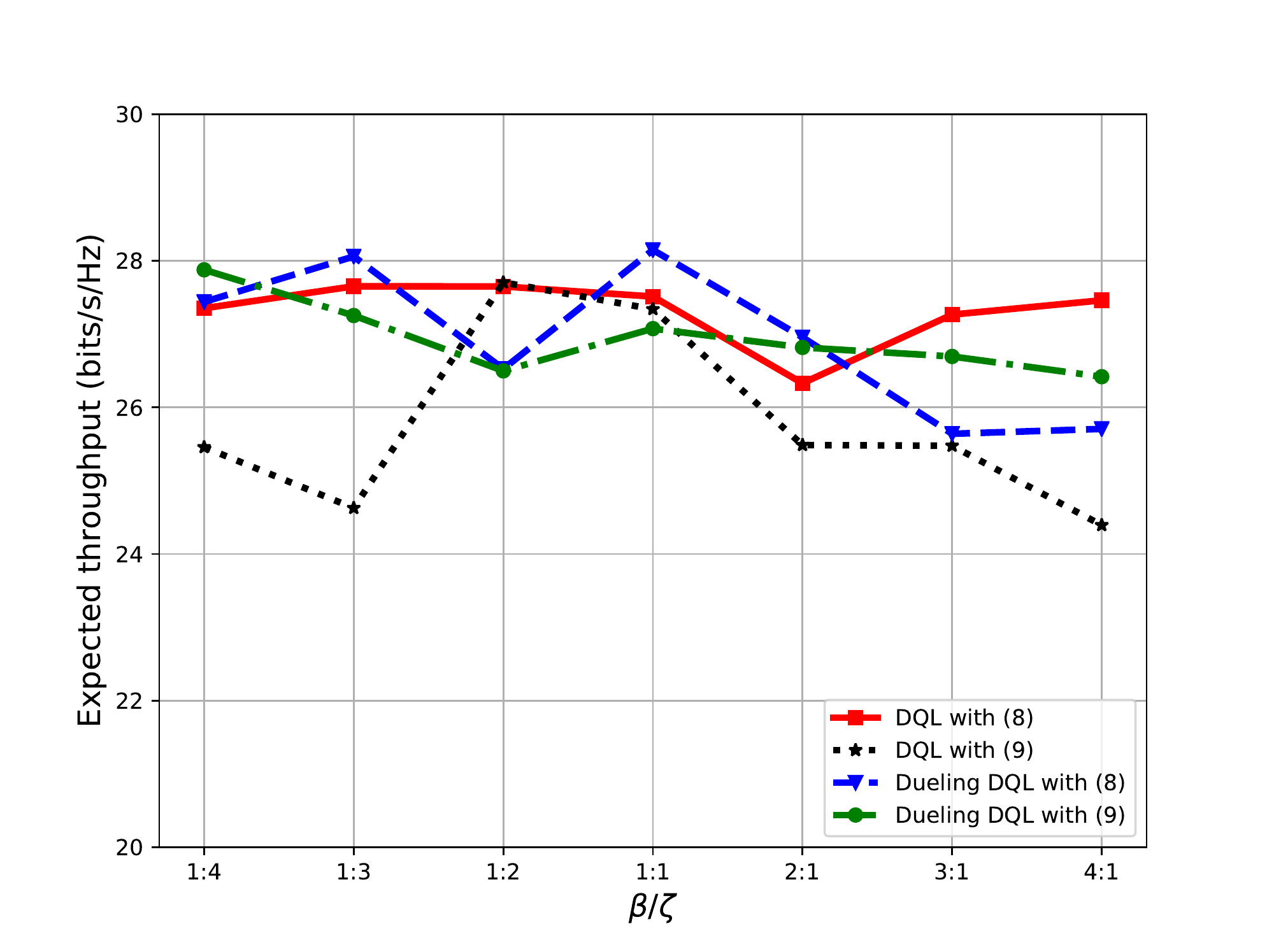}}
	\caption{The network's sum-rate when using the DQL and dueling DQL algorithms with 3 clusters}
	\label{fig:throughput3clusters}
\end{figure}

In (\ref{equ:reward}), we consider two elements: the trajectory cost and the average throughput. In order to quantify the communication efficiency, we compare the total throughput in different scenarios. In Fig. (\ref{fig:throughput3clusters}), the performances of the DQL algorithm associated with several $\beta/\zeta$ values are considered while using the binary trajectory function (\ref{equ:Rplus1}), the episode reward (\ref{equ:R2}) and $3$ clusters. The throughput obtained for $\beta/\zeta= 1:1$ is higher than that of the others and when $\beta$ increases, the performance degrades. However, when comparing with the Fig. (\ref{fig:reward3clusters}b), we realise that in some scenarios the UAV was stuck and could not find the way to the destination. That leads to increased flight time spent and distance travelled. More details are shown in Fig. (\ref{fig:throughput3clusters}b), where we compare the expected throughput of both the DQL and dueling DQL algorithms. The best throughput is achieved when using the dueling DQL algorithm with $\beta/\zeta= 1:1$ in conjunction with (\ref{equ:Rplus1}), which is higher than the peak of the DQL method with $\beta/\zeta= 1:2$.

\begin{figure}[h!]
	\centering
	\subfigure[With (\ref{equ:Rplus1}), (\ref{equ:R2})]{\includegraphics[width=0.65\textwidth]{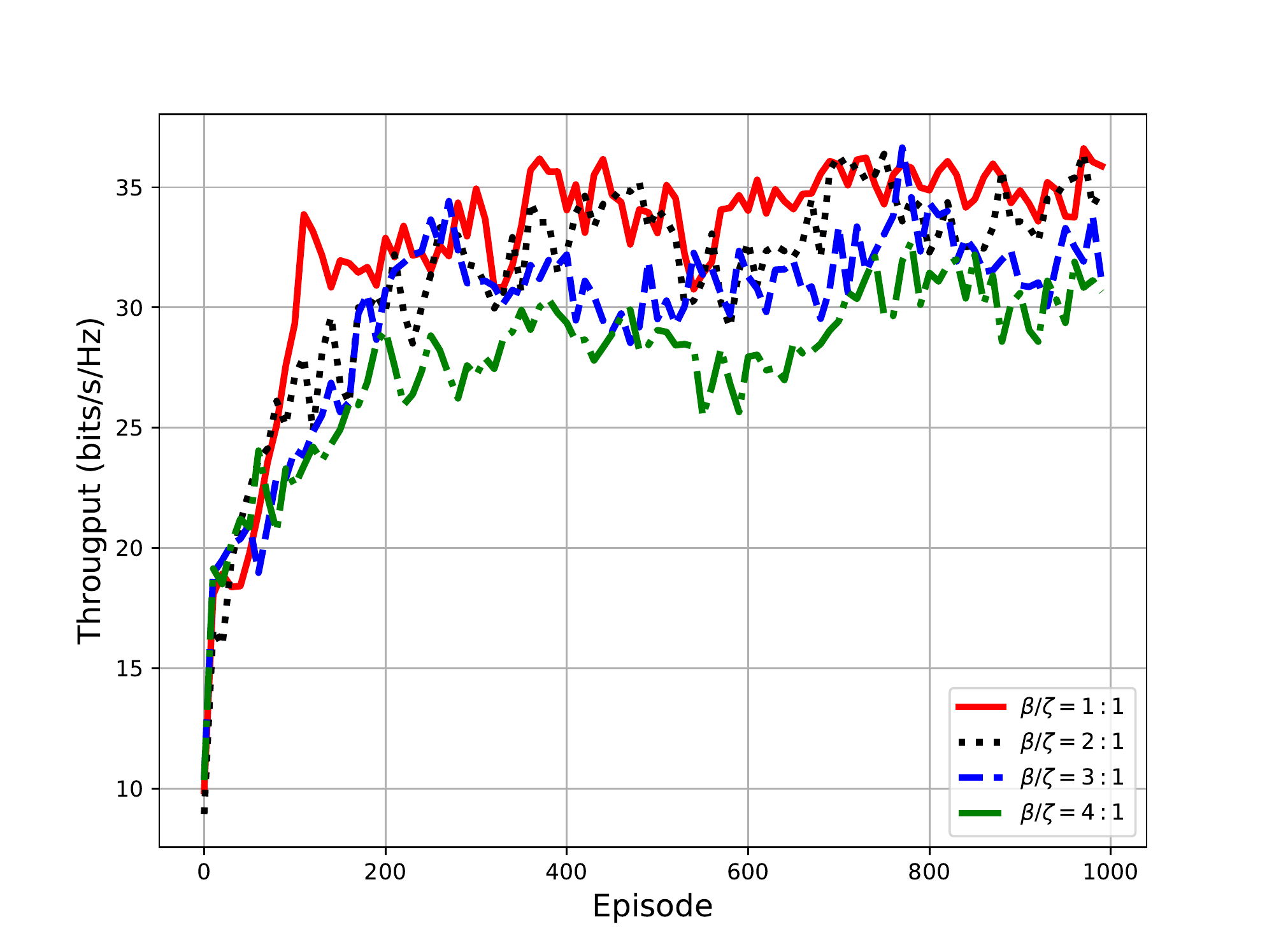}}
	\subfigure[With  (\ref{equ:Rplus2}), (\ref{equ:R2})]{\includegraphics[width=0.65\textwidth]{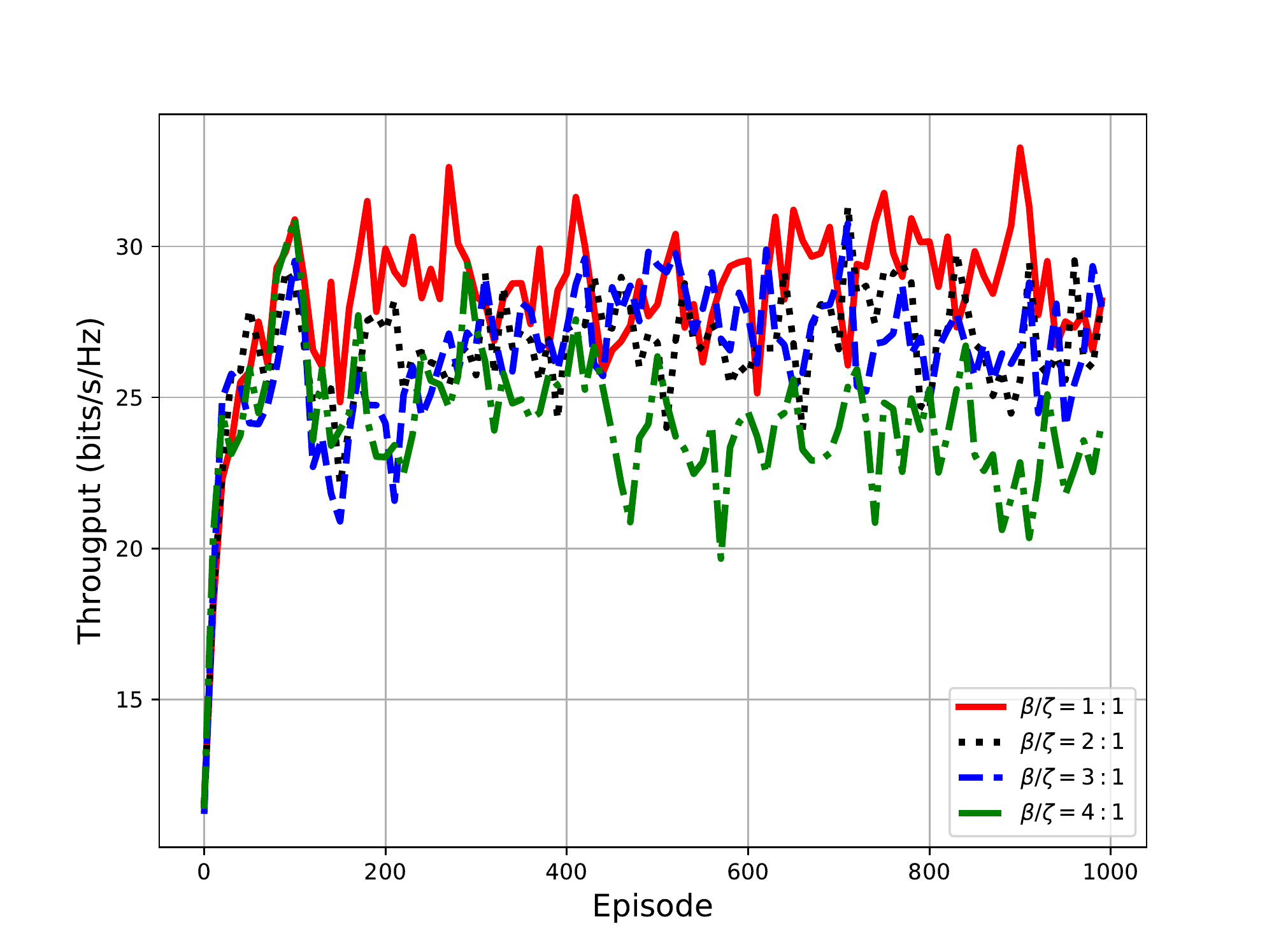}}
	\caption{The obtained total throughput when using the DQL algorithm with 5 clusters}
	\label{fig:throughput5clusters}
\end{figure}

In Fig. (\ref{fig:throughput5clusters}), we compare the throughput of different techniques in the $5$-cluster scenario. Let us now consider the binary trajectory design function (\ref{equ:Rplus1}) in Fig. (\ref{fig:throughput5clusters}a),  where the DQL algorithm achieves the best performance using $\beta/\zeta= 1:1$ and $\beta/\zeta = 2:1$. There is a slight difference between the DQL method having different settings, when using exponential the trajectory design function (\ref{equ:Rplus2}), as shown in Fig. (\ref{fig:throughput5clusters}b).

\begin{figure}[h!]
	\centering
	\subfigure[]{\includegraphics[width=0.65\textwidth]{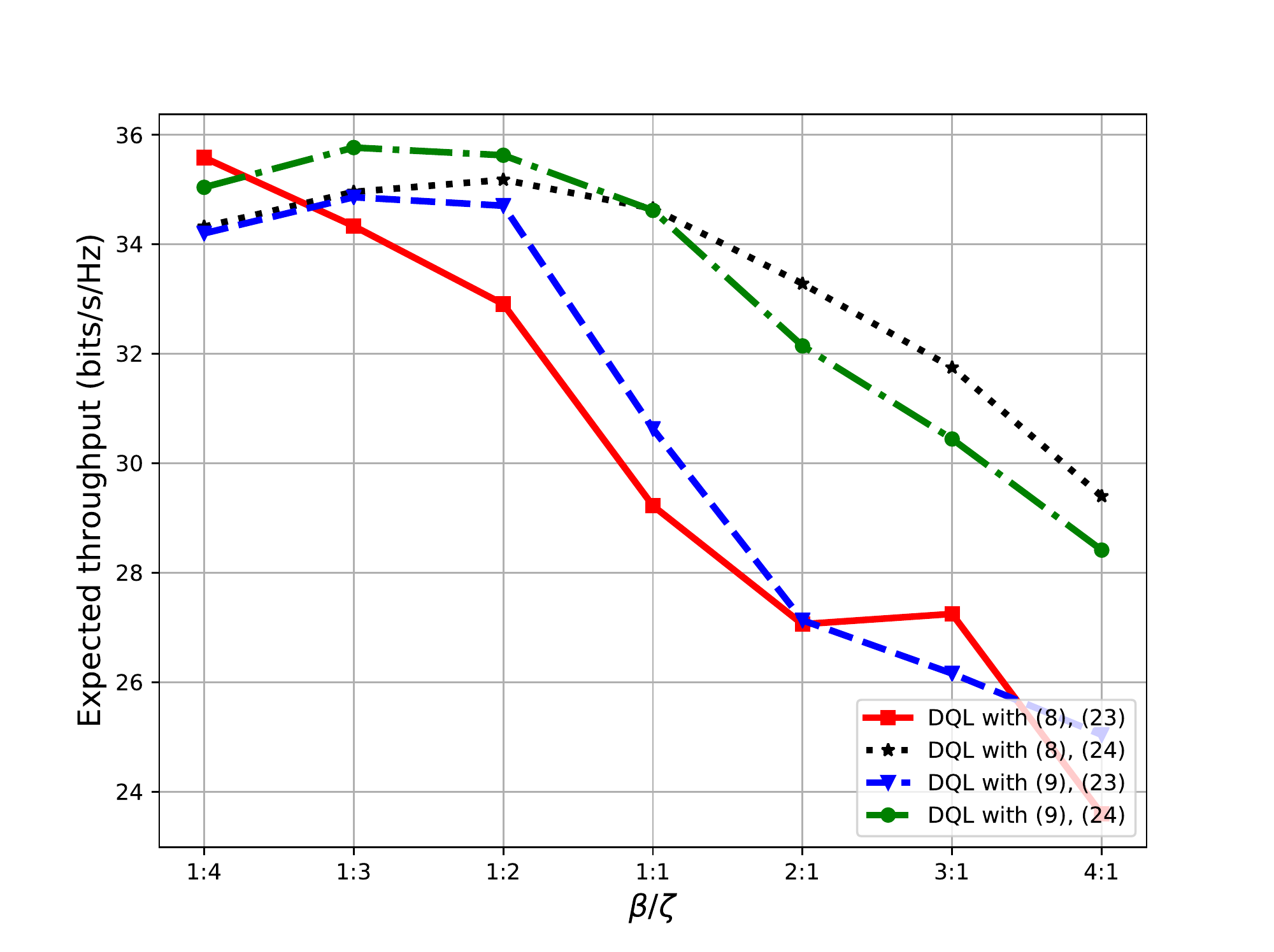}}
	\subfigure[]{\includegraphics[width=0.65\textwidth]{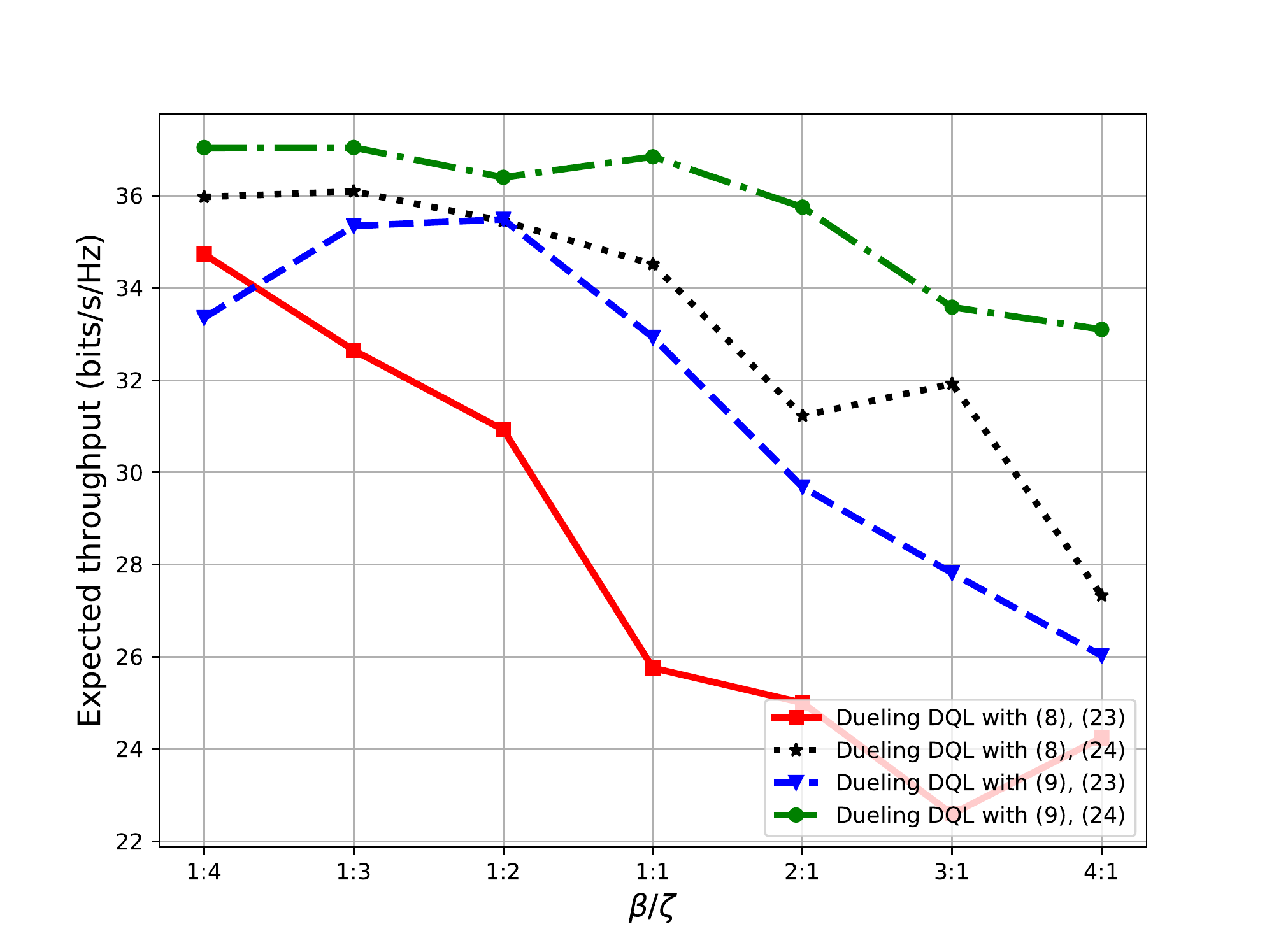}}
	\caption{The obtained throughput when using the DQL and dueling DQL algorithms in 5-cluster scenario}
	\label{fig:throughput5clusters2}
\end{figure}

\begin{figure}[h!]
	\centering
	\subfigure[With (\ref{equ:Rplus2})]{\includegraphics[width=0.65\textwidth]{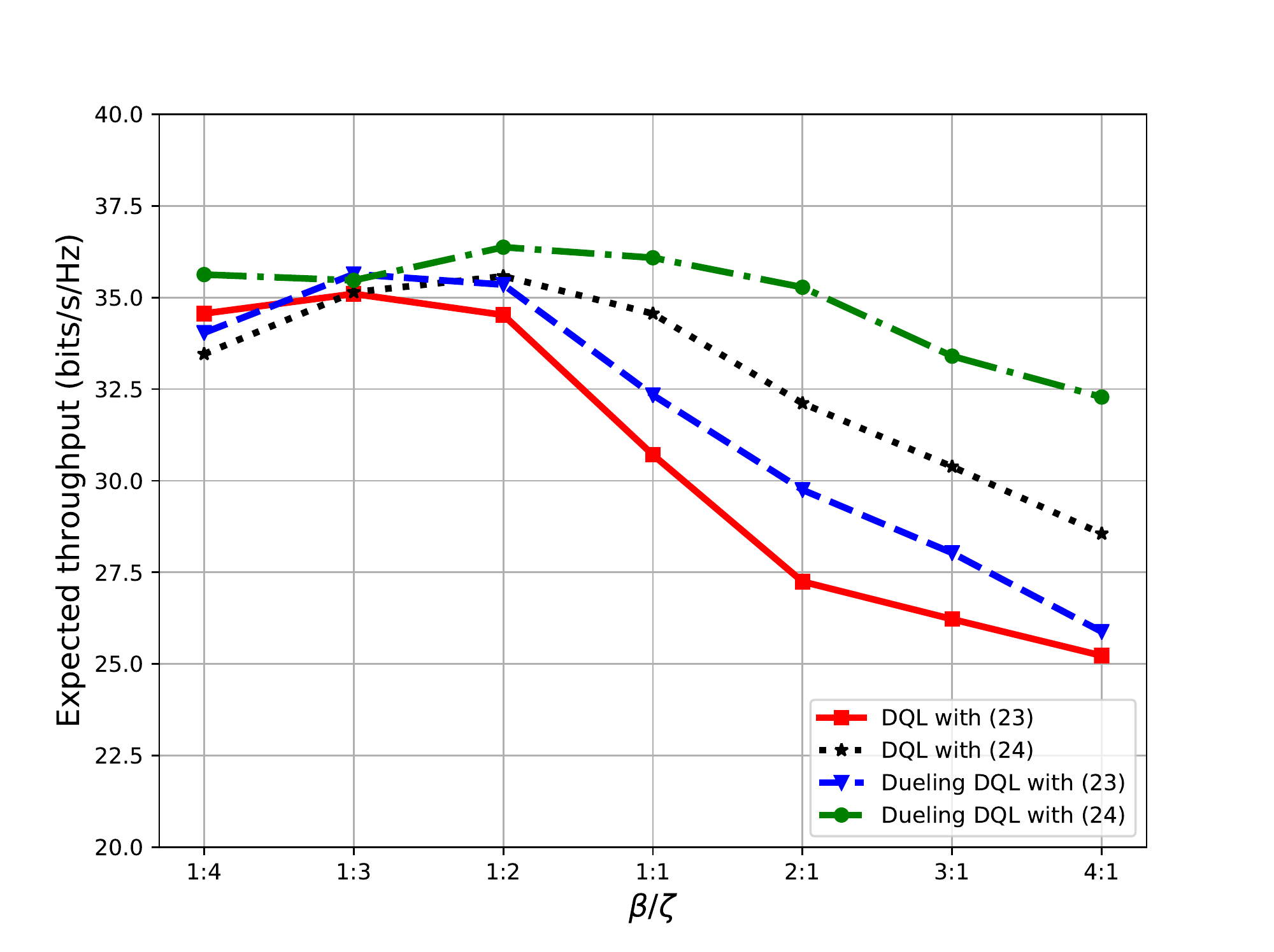}}
	\subfigure[With (\ref{equ:R2})]{\includegraphics[width=0.65\textwidth]{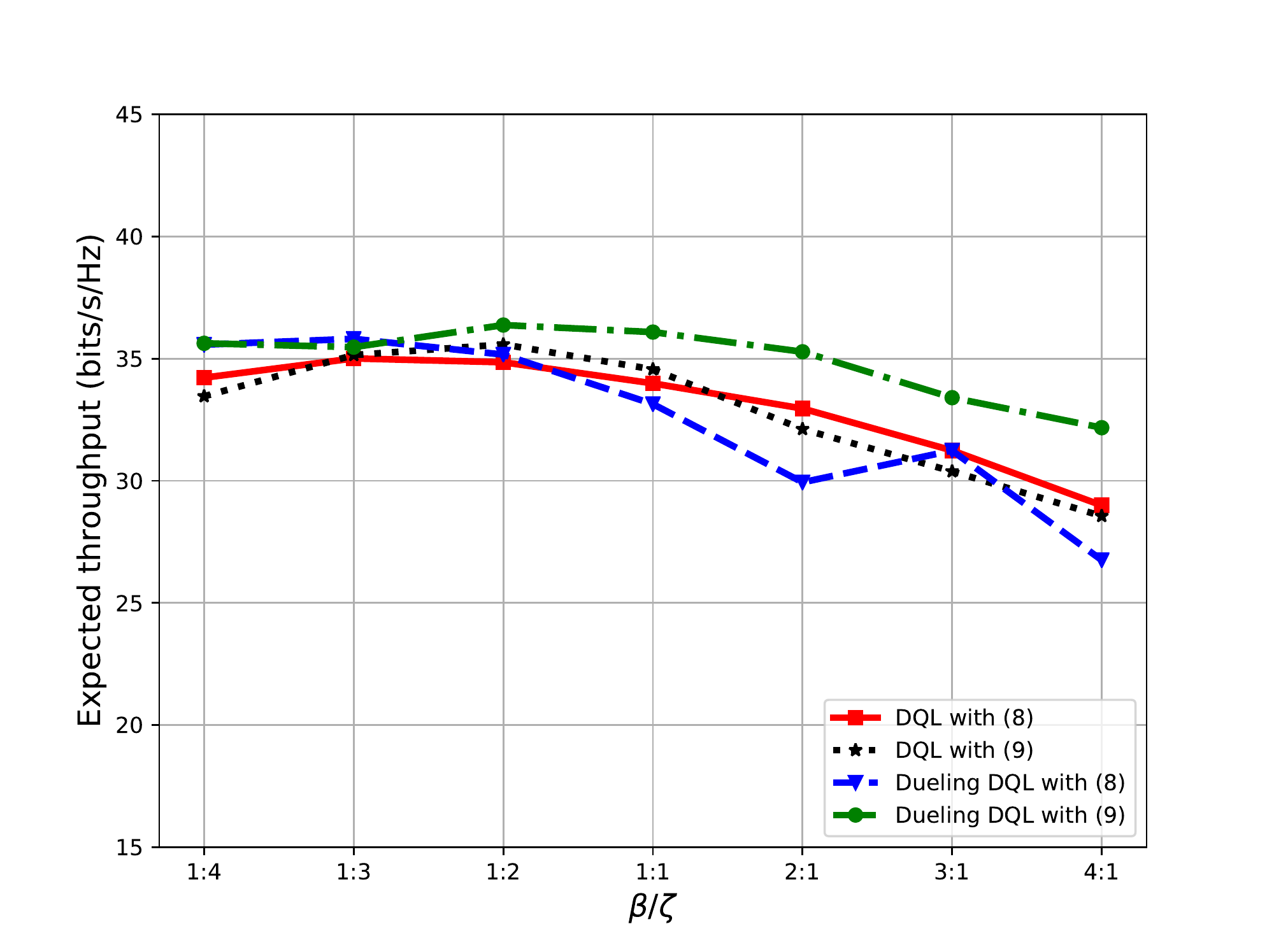}}
	\caption{The expected throughput when using the DQL and dueling DQL algorithms with 5 clusters}
	\label{fig:throughput5clusters3}
\end{figure}

In Fig. (\ref{fig:throughput5clusters2}) and Fig. (\ref{fig:throughput5clusters3}), we compare the throughput of different $\beta/\zeta$ pairs. The DQL algorithm reaches the optimal throughput with the aid of trial-and-learn methods, hence it is important to carefully design the reward function to avoid excessive offline training. As shown in Fig. (\ref{fig:throughput5clusters2}), the DQL and dueling DQL algorithm exhibit reasonable stability for several $\beta/\zeta \le 1:1$ pairs as well as reward functions. While we can achieve the similar expected reward with different reward setting in Fig. (\ref{fig:reward5clusters3}), the throughput is degraded when the $\beta/\zeta$ increases. In contrast, with higher $\beta$ values, the UAV can finish the mission faster. It is a trade-off game when we can choose an approximate $\beta/\zeta$ value for our specific purposes. When we employ the DQL and the dueling DQL algorithms with the episode reward (\ref{equ:R2}), the throughput attained is higher than that using the immediate reward (\ref{equ:R1}) with different $\beta/\zeta$ values.

Furthermore, we compare the expected throughput of the DQL and of the dueling DQL algorithm when using the  exponential trajectory design (\ref{equ:Rplus2}) in Fig. (\ref{fig:throughput5clusters3}a) and the episode reward (\ref{equ:R2}) in Fig. (\ref{fig:throughput5clusters3}b). In Fig. (\ref{fig:throughput5clusters3}a), the dueling DQL method outperforms the DQL algorithm for almost all $\beta/\zeta$ values in both function (\ref{equ:R1}) and (\ref{equ:R2}). When we use the episode reward (\ref{equ:R2}), the obtained throughput are stable with different $\beta/\zeta$ values. The throughput attained by using the exponential function (\ref{equ:Rplus2}) is higher than that using the binary trajectory (\ref{equ:Rplus1}) and by using the episode reward (\ref{equ:R2}) is higher than that using the immediate reward (\ref{equ:R1}). We can achieve the best performance when using the dueling DQL algorithm with (\ref{equ:Rplus2}) and (\ref{equ:R2}). However, in some scenarios, we can achieve the better performance with different algorithmic setting as we can see in Fig. (\ref{fig:throughput3clusters}b) and Fig. (\ref{fig:throughput5clusters2}a). Thus, there is a trade-off governing the choice of the algorithm and function design.

\subsection{Parametric Study}
\begin{figure}[t!]
	\centering
	\subfigure{\includegraphics[width=0.65\textwidth]{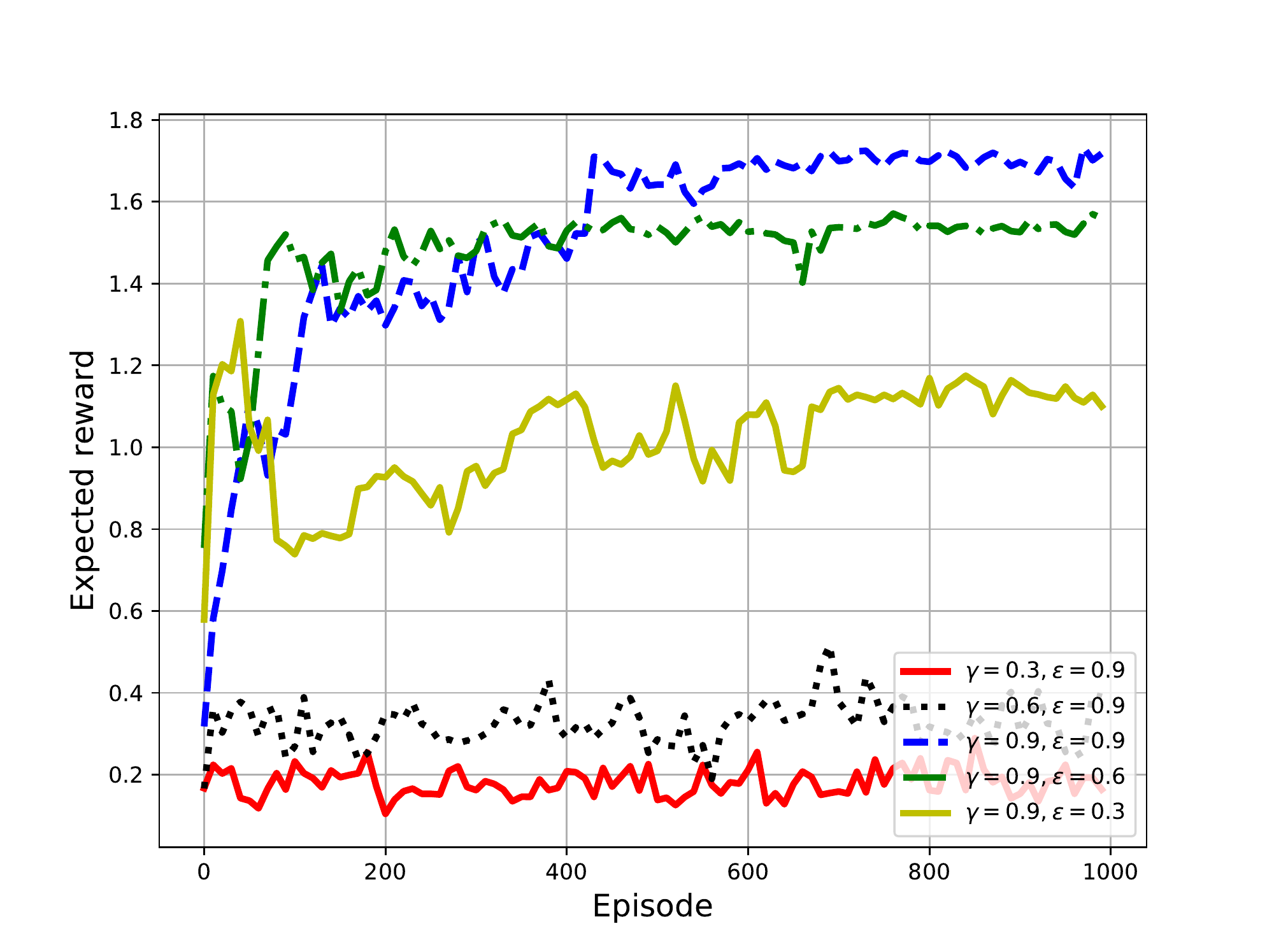}}
	\caption{The performance when using the DQL algorithm with different discount factors, $\gamma$, and exploration factors, $\epsilon$}
	\label{fig:exp}
\end{figure}

In Fig. (\ref{fig:exp}), we compare the performance of our DQL technique using different $\textit{exploration}$ parameters $\gamma$ and $\epsilon$ values in our $\epsilon$-greedy method. The DQL algorithm achieves the best performance with the discounting factor of $\gamma = 0.9$ and $\epsilon = 0.9$ in the $5$-cluster scenario of Fig.~(\ref{fig:exp}). Balancing the \textit{exploration} and \textit{exploitation} as well as the action chosen is quite challenging, in order to maintain a steady performance of the DQL algorithm. Based on the results of Fig. (\ref{fig:exp}), we opted for $\gamma = 0.9$ and $\epsilon = 0.9$ for our algorithmic setting.

\begin{figure}[h!]
	\centering
	\subfigure{\includegraphics[width=0.65\textwidth]{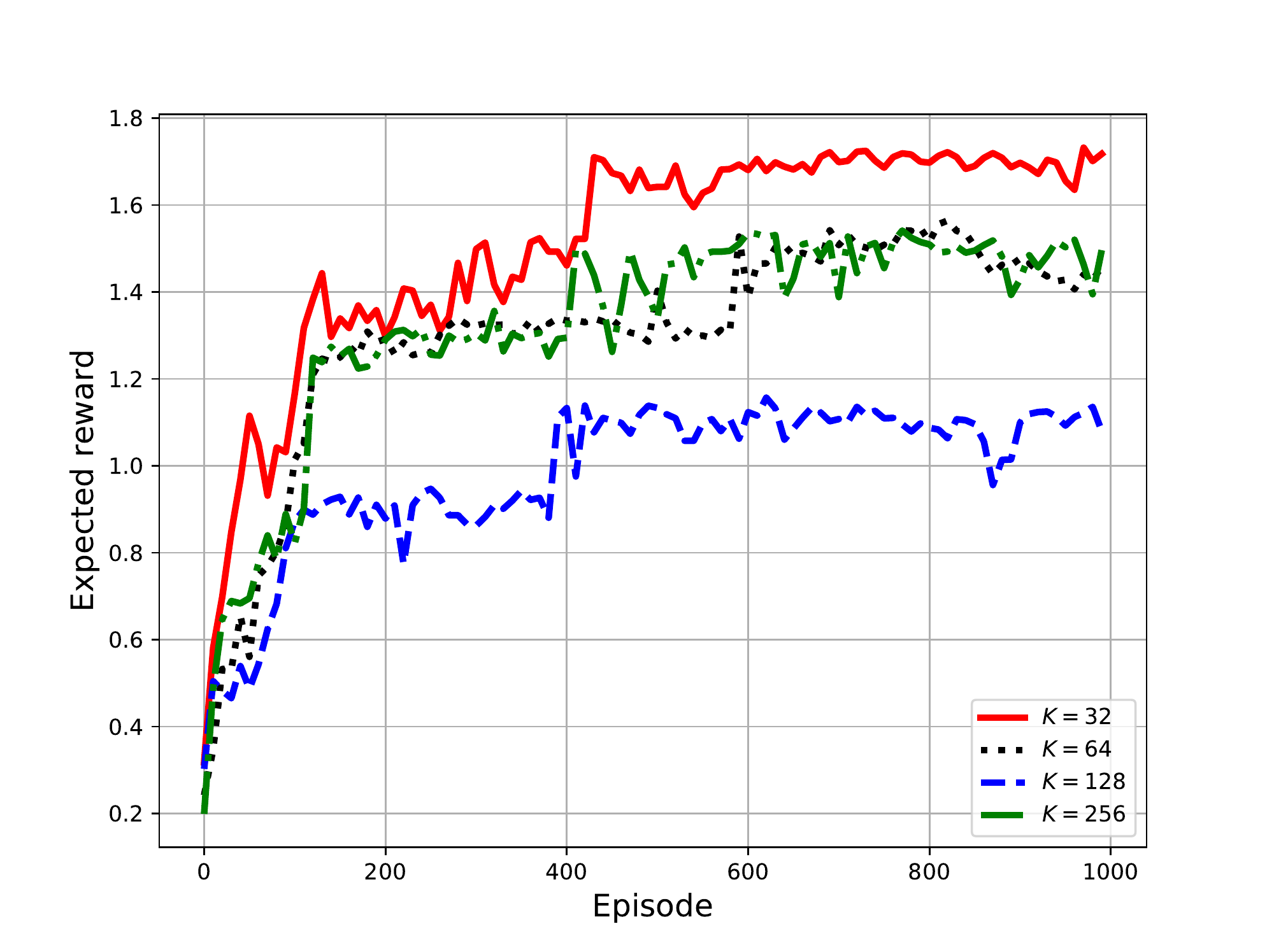}}
	\caption{The performance when using the DQL algorithm in $5$-cluster scenario and different batch sizes, $K$}
	\label{fig:batch3clusters}
\end{figure}	
%

Next, we compare the expected reward of different mini-batch sizes, $K$. In the $5$-cluster scenario of Fig. (\ref{fig:batch3clusters}), the DQL achieves the optimal performance with a batch size of $K = 32$. There is a slight difference in terms of convergence speed with batch size $K = 32$ is the fastest. Overall, we set the mini-batch size to $K = 32$ for our DQL algorithm.

\begin{figure}[h!]
	\centering
	\subfigure{\includegraphics[width=0.65\textwidth]{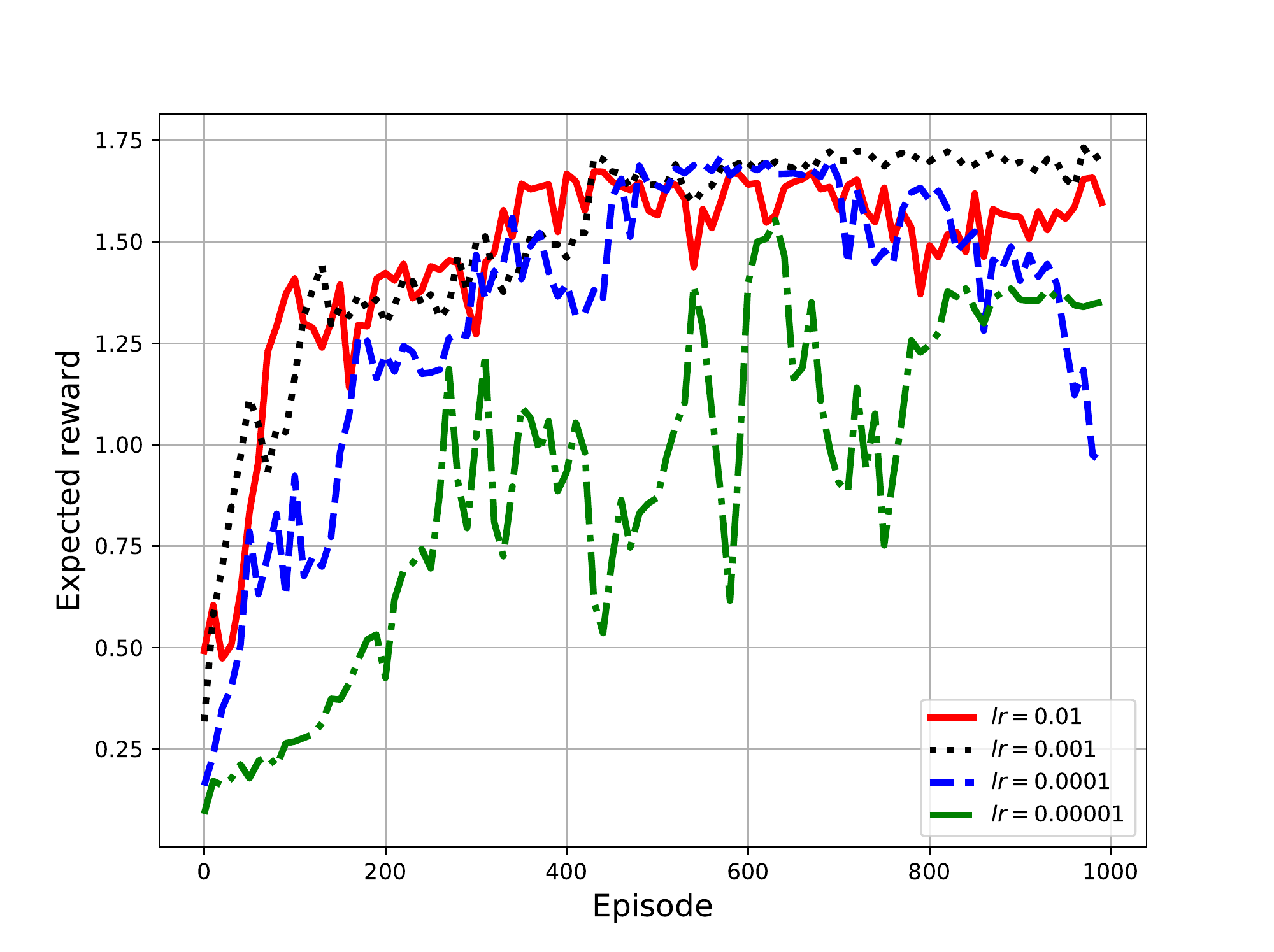}}
	\caption{The performance when using DQL algorithm with different learning rate, $lr$}
	\label{fig:lr}
\end{figure}

Fig. (\ref{fig:lr}) shows the performance of the DQL algorithm with different learning rates in updating the neural networks parameters while considering the scenarios of $5$ clusters. When the learning rate is as high as $\alpha = 0.01$, the pace of updating the network may result the fluctuating performance. Moreover, when $\alpha = 0.0001$ or $\alpha = 0.00001$ the convergence speed is slower and may be stuck in a local optimum instead reaching the global optimum. Thus, based on our experiments, we opted for the learning rate of $\alpha = 0.001$ for the algorithms.


\section{Conclusion}\label{Sec:Con}
In this paper, the DRL technique has been proposed jointly optimising the flight trajectory and data collection performance of UAV-assisted IoT networks. The optimisation game has been formulated to balance the flight time and total throughput while guaranteeing the quality-of-service constraints. Bearing in mind the limited UAV power level and the associated communication constraints, we proposed a DRL technique for maximising the throughput while the UAV has to move along the shortest path to reach the destination. Both the DQL and dueling DQL techniques having a low computational complexity have been conceived. Our simulation results showed the efficiency of our techniques both in simple and complex environmental settings.


\bibliographystyle{IEEEtran}
\bibliography{IEEEabrv,reference}

\begin{thebibliography}{10}
\providecommand{\url}[1]{#1}
\csname url@samestyle\endcsname
\providecommand{\newblock}{\relax}
\providecommand{\bibinfo}[2]{#2}
\providecommand{\BIBentrySTDinterwordspacing}{\spaceskip=0pt\relax}
\providecommand{\BIBentryALTinterwordstretchfactor}{4}
\providecommand{\BIBentryALTinterwordspacing}{\spaceskip=\fontdimen2\font plus
\BIBentryALTinterwordstretchfactor\fontdimen3\font minus
  \fontdimen4\font\relax}
\providecommand{\BIBforeignlanguage}[2]{{%
\expandafter\ifx\csname l@#1\endcsname\relax
\typeout{** WARNING: IEEEtran.bst: No hyphenation pattern has been}%
\typeout{** loaded for the language `#1'. Using the pattern for}%
\typeout{** the default language instead.}%
\else
\language=\csname l@#1\endcsname
\fi
#2}}
\providecommand{\BIBdecl}{\relax}
\BIBdecl

\bibitem{Drone:UK}
\BIBentryALTinterwordspacing
``Drone trial to help {Isle of Wight} receive medical supplies faster during
  {COVID19} pandemic.'' [Online]. Available:
  \url{https://www.southampton.ac.uk/news/2020/04/drones-covid-iow.page}
\BIBentrySTDinterwordspacing

\bibitem{Drone:Chile}
\BIBentryALTinterwordspacing
``This {Chilean} community is using drones to deliver medicine to the
  elderly.'' [Online]. Available:
  \url{https://www.weforum.org/agenda/2020/04/drone-chile-covid19/}
\BIBentrySTDinterwordspacing

\bibitem{MG:17:SC}
M.~Gao, X.~Xu, Y.~Klinger, J.~van~der Woerd, and P.~Tapponnier,
  ``High-resolution mapping based on an unmanned aerial vehicle {(UAV)} to
  capture paleoseismic offsets along the {Altyn-Tagh} fault, {China},''
  \emph{Sci. Rep.}, vol.~7, no.~1, pp. 1--11, Aug. 2017.

\bibitem{QL:16:VTM}
Q.~Liu, J.~Wu, P.~Xia, S.~Zhao, Y.~Yang, W.~Chen, and L.~Hanzo, ``Charging
  unplugged: Will distributed laser charging for mobile wireless power transfer
  work?'' \emph{IEEE Vehicular Technology Magazine}, vol.~11, no.~4, pp.
  36--45, Dec. 2016.

\bibitem{HC:06:ICC}
H.~Claussen, ``Distributed algorithms for robust self-deployment and load
  balancing in autonomous wireless access networks,'' in \emph{Proc. IEEE Int.
  Conf. on Commun. (ICC)}, vol.~4, Istanbul, Turkey, June 2006, pp. 1927--1932.

\bibitem{JG:18:SAC}
J.~Gong, T.-H. Chang, C.~Shen, and X.~Chen, ``Flight time minimization of {UAV}
  for data collection over wireless sensor networks,'' \emph{IEEE J. Select.
  Areas Commun.}, vol.~36, no.~9, pp. 1942--1954, Sept. 2018.

\bibitem{CZ:20:CCN}
C.~Zhong, M.~C. Gursoy, and S.~Velipasalar, ``Deep reinforcement learning-based
  edge caching in wireless networks,'' \emph{IEEE Trans. Cogn. Commun. Netw.},
  vol.~6, no.~1, pp. 48--61, Mar. 2020.

\bibitem{HW:20:WC}
H.~Wu, Z.~Wei, Y.~Hou, N.~Zhang, and X.~Tao, ``Cell-edge user offloading via
  flying {UAV} in non-uniform heterogeneous cellular networks,'' \emph{IEEE
  Trans. Wireless Commun.}, vol.~19, no.~4, pp. 2411--2426, Apr. 2020.

\bibitem{HH:20:VT}
H.~Huang \emph{et~al.}, ``Deep reinforcement learning for {UAV} navigation
  through massive {MIMO} technique,'' \emph{{IEEE} Trans. Veh. Technol.},
  vol.~69, no.~1, pp. 1117--1121, Jan. 2020.

\bibitem{TD:19:GLOBECOM}
T.~Q. Duong, L.~D. Nguyen, H.~D. Tuan, and L.~Hanzo, ``Learning-aided realtime
  performance optimisation of cognitive {UAV}-assisted disaster
  communication,'' in \emph{Proc. IEEE Global Communications Conference
  (GLOBECOM)}, Waikoloa, HI, USA, Dec. 2019.

\bibitem{Trung:19:IWCMC}
T.~Q. Duong, L.~D. Nguyen, and L.~K. Nguyen, ``Practical optimisation of path
  planning and completion time of data collection for {UAV}-enabled disaster
  communications,'' in \emph{Proc. 15th Int. Wireless Commun. Mobile Computing
  Conf. (IWCMC)}, Tangier, Morocco, Jun. 2019, pp. 372--377.

\bibitem{MM:16:WC}
M.~Mozaffari, W.~Saad, M.~Bennis, and M.~Debbah, ``Unmanned aerial vehicle with
  underlaid device-to-device communications: Performance and tradeoffs,''
  \emph{IEEE Trans. Wireless Commun.}, vol.~15, no.~6, pp. 3949--3963, Jun.
  2016.

\bibitem{Long:EAI}
L.~D. Nguyen, A.~Kortun, and T.~Q. Duong, ``An introduction of real-time
  embedded optimisation programming for {UAV} systems under disaster
  communication,'' \emph{{EAI} Endorsed Transactions on Industrial Networks and
  Intelligent Systems}, vol.~5, no.~17, pp. 1--8, Dec. 2018.

\bibitem{Minh:19:WCL}
M.-N. Nguyen, L.~D. Nguyen, T.~Q. Duong, and H.~D. Tuan, ``Real-time optimal
  resource allocation for embedded {UAV} communication systems,'' \emph{IEEE
  Wireless Commun. Lett.}, vol.~8, no.~1, pp. 225--228, Feb. 2019.

\bibitem{Xiaowei:19:VT}
X.~Li, H.~Yao, J.~Wang, X.~Xu, C.~Jiang, and L.~Hanzo, ``A near-optimal
  {UAV}-aided radio coverage strategy for dense urban areas,'' \emph{{IEEE}
  Trans. Veh. Technol.}, vol.~68, no.~9, pp. 9098--9109, Sept. 2019.

\bibitem{HZ:20:VT}
H.~Zhang and L.~Hanzo, ``Federated learning assisted multi-{UAV} networks,''
  \emph{{IEEE} Trans. Veh. Technol.}, vol.~69, no.~11, pp. 14\,104--14\,109,
  Nov. 2020.

\bibitem{Xiao:19:VT}
X.~Liu, Y.~Liu, Y.~Chen, and L.~Hanzo, ``Trajectory design and power control
  for multi-{UAV} assisted wireless networks: A machine learning approach,''
  \emph{{IEEE} Trans. Veh. Technol.}, vol.~68, no.~8, pp. 7957--7969, Aug.
  2019.

\bibitem{KK:19:Access}
K.~K. Nguyen, T.~Q. Duong, N.~A. Vien, N.-A. Le-Khac, and L.~D. Nguyen,
  ``Distributed deep deterministic policy gradient for power allocation control
  in {D2D}-based {V2V} communications,'' \emph{IEEE Access}, vol.~7, pp.
  164\,533--164\,543, Nov. 2019.

\bibitem{Khoi:19:Access}
K.~K. Nguyen, T.~Q. Duong, N.~A. Vien, N.-A. Le-Khac, and N.~M. Nguyen,
  ``Non-cooperative energy efficient power allocation game in {D2D}
  communication: A multi-agent deep reinforcement learning approach,''
  \emph{IEEE Access}, vol.~7, pp. 100\,480--100\,490, Jul. 2019.

\bibitem{Khoi:20:Access}
K.~K. Nguyen, N.~A. Vien, L.~D. Nguyen, M.-T. Le, L.~Hanzo, and T.~Q. Duong,
  ``Real-time energy harvesting aided scheduling in {UAV}-assisted {D2D}
  networks relying on deep reinforcement learning,'' \emph{IEEE Access},
  vol.~9, pp. 3638--3648, 2021.

\bibitem{KL:19:VT}
K.~Li, W.~Ni, E.~Tovar, and A.~Jamalipour, ``On-board deep {Q}-network for
  {UAV}-assisted online power transfer and data collection,'' \emph{{IEEE}
  Trans. Veh. Technol.}, vol.~68, no.~12, pp. 12\,215--12\,226, Dec. 2019.

\bibitem{UC:19:WC}
U.~Challita, W.~Saad, and C.~Bettstetter, ``Interference management for
  cellular-connected {UAVs}: A deep reinforcement learning approach,''
  \emph{IEEE Trans. Wireless Commun.}, vol.~18, no.~4, pp. 2125--2140, Apr.
  2019.

\bibitem{XL:19:VT}
X.~Liu, Y.~Liu, and Y.~Chen, ``Reinforcement learning in multiple-{UAV}
  networks: Deployment and movement design,'' \emph{{IEEE} Trans. Veh.
  Technol.}, vol.~68, no.~8, pp. 8036--8049, Aug. 2019.

\bibitem{CW:19:VT}
C.~Wang, J.~Wang, Y.~Shen, and X.~Zhang, ``Autonomous navigation of {UAVs} in
  large-scale complex environments: A deep reinforcement learning approach,''
  \emph{{IEEE} Trans. Veh. Technol.}, vol.~68, no.~3, pp. 2124--2136, Mar.
  2019.

\bibitem{SG:17:ICRA}
S.~Gu, E.~Holly, T.~Lillicrap, and S.~Levine, ``Deep reinforcement learning for
  robotic manipulation with asynchronous off-policy updates,'' in \emph{Proc.
  IEEE International Conf. Robot. Autom. (ICRA)}, May 2017, pp. 3389--3396.

\bibitem{QC:18:AAAI}
Q.~Cai, A.~Filos-Ratsikas, P.~Tang, and Y.~Zhang, ``Reinforcement mechanism
  design for fraudulent behaviour in e-commerce,'' in \emph{Thirty-Second AAAI
  Conf. Artif. Intell.}, 2018.

\bibitem{Mnih:13}
\BIBentryALTinterwordspacing
V.~Mnih, K.~Kavukcuoglu, D.~Silver, A.~Graves, I.~Antonoglou, D.~Wierstra, and
  M.~Riedmiller, ``Playing {Atari} with deep reinforcement learning,'' 2013.
  [Online]. Available: \url{arXiv preprint arXiv:1312.5602}
\BIBentrySTDinterwordspacing

\bibitem{YY:19:SAC}
Y.~Yu, T.~Wang, and S.~C. Liew, ``Deep-reinforcement learning multiple access
  for heterogeneous wireless networks,'' \emph{IEEE J. Select. Areas Commun.},
  vol.~37, no.~6, pp. 1277--1290, Jun. 2019.

\bibitem{NZ:19:WC}
N.~Zhao, Y.-C. Liang, D.~Niyato, Y.~Pei, M.~Wu, and Y.~Jiang, ``Deep
  reinforcement learning for user association and resource allocation in
  heterogeneous cellular networks,'' \emph{IEEE Trans. Wireless Commun.},
  vol.~18, no.~11, pp. 5141--5152, Nov. 2019.

\bibitem{SY:19:VT}
S.~Yin, S.~Zhao, Y.~Zhao, , and F.~R. Yu, ``Intelligent trajectory design in
  {UAV}-aided communications with reinforcement learning,'' \emph{{IEEE} Trans.
  Veh. Technol.}, vol.~68, no.~8, pp. 8227--8231, Aug. 2019.

\bibitem{DY:18:VT}
D.~Yang, Q.~Wu, Y.~Zeng, and R.~Zhang, ``Energy tradeoff in ground-to-{UAV}
  communication via trajectory design,'' \emph{{IEEE} Trans. Veh. Technol.},
  vol.~67, no.~7, pp. 6721--6726, Jul. 2018.

\bibitem{HW:19:WC}
H.~Wang, J.~Wang, G.~Ding, J.~Chen, F.~Gao, and Z.~Han, ``Completion time
  minimization with path planning for fixed-wing {UAV} communications,''
  \emph{IEEE Trans. Wireless Commun.}, vol.~18, no.~7, pp. 3485--3499, Jul.
  2019.

\bibitem{Huy:20:CCN}
H.~T. Nguyen, H.~D. Tuan, T.~Q. Duong, H.~V. Poor, and W.-J. Hwang, ``Joint
  {D2D} assignment, bandwidth and power allocation in cognitive {UAV}-enabled
  networks,'' \emph{IEEE Trans. Cogn. Commun. Netw.}, vol.~6, no.~3, pp.
  1084--1095, Sept. 2020.

\bibitem{LL:19:WC}
L.~Liu, S.~Zhang, and R.~Zhang, ``Multi-beam {UAV} communication in cellular
  uplink: Cooperative interference cancellation and sum-rate maximization,''
  \emph{IEEE Trans. Wireless Commun.}, vol.~18, no.~10, pp. 4679--4691, Oct.
  2019.

\bibitem{LX:19:IOT}
L.~Xie, J.~Xu, and R.~Zhang, ``Throughput maximization for {UAV}-enabled
  wireless powered communication networks,'' \emph{{IEEE} Internet Things J.},
  vol.~6, no.~2, pp. 1690--1703, Apr. 2019.

\bibitem{LN:19:SPAWC}
L.~D. Nguyen, K.~K. Nguyen, A.~Kortun, and T.~Q. Duong, ``Real-time deployment
  and resource allocation for distributed {UAV} systems in disaster relief,''
  in \emph{Proc. IEEE 20th International Workshop on Signal Processing Advances
  in Wireless Commun. (SPAWC)}, Cannes, France, Jul. 2019, pp. 1--5.

\bibitem{QW:18:WC}
Q.~Wu, Y.~Zeng, and R.~Zhang, ``Joint trajectory and communication design for
  multi-{UAV} enabled wireless networks,'' \emph{IEEE Trans. Wireless Commun.},
  vol.~17, no.~3, pp. 2109--2121, Mar. 2018.

\bibitem{CZ:18:WCL}
C.~Zhan, Y.~Zeng, and R.~Zhang, ``Energy-efficient data collection in {UAV}
  enabled wireless sensor network,'' \emph{IEEE Wireless Commun. Lett.},
  vol.~7, no.~3, pp. 328--331, Jun. 2018.

\bibitem{HW:18:WCL}
H.~Wang, G.~Ren, J.~Chen, G.~Ding, and Y.~Yang, ``Unmanned aerial vehicle-aided
  communications: Joint transmit power and trajectory optimization,''
  \emph{IEEE Wireless Commun. Lett.}, vol.~7, no.~4, pp. 522--525, Aug. 2018.

\bibitem{ZW:20:ITJ}
Z.~Wang, R.~Liu, Q.~Liu, J.~S. Thompson, and M.~Kadoch, ``Energy-efficient data
  collection and device positioning in {UAV}-assisted {IoT},'' \emph{{IEEE}
  Internet Things J.}, vol.~7, no.~2, pp. 1122--1139, Feb. 2020.

\bibitem{JL:20:ITJ}
J.~Li \emph{et~al.}, ``Joint optimization on trajectory, altitude, velocity,
  and link scheduling for minimum mission time in {UAV}-aided data
  collection,'' \emph{{IEEE} Internet Things J.}, vol.~7, no.~2, pp.
  1464--1475, Feb. 2020.

\bibitem{MS:20:WC}
M.~Samir, S.~Sharafeddine, C.~M. Assi, T.~M. Nguyen, and A.~Ghrayeb, ``{UAV}
  trajectory planning for data collection from time-constrained {IoT}
  devices,'' \emph{IEEE Trans. Wireless Commun.}, vol.~19, no.~1, pp. 34--46,
  Jan. 2020.

\bibitem{MH:20:C}
M.~Hua, L.~Yang, Q.~Wu, and A.~L. Swindlehurst, ``{3D UAV} trajectory and
  communication design for simultaneous uplink and downlink transmission,''
  \emph{IEEE Trans. on Commun.}, vol.~68, no.~9, pp. 5908--5923, 2020.

\bibitem{CZ:20:C}
C.~Zhan and Y.~Zeng, ``Aerial–ground cost tradeoff for multi-{UAV}-enabled
  data collection in wireless sensor networks,'' \emph{IEEE Trans. on Commun.},
  vol.~68, no.~3, pp. 1937--1950, March 2020.

\bibitem{MS:20:VT}
M.~Samir, C.~Assi, S.~Sharafeddine, D.~Ebrahimi, and A.~Ghrayeb, ``Age of
  information aware trajectory planning of {UAVs} in intelligent transportation
  systems: A deep learning approach,'' \emph{{IEEE} Trans. Veh. Technol.},
  vol.~69, no.~11, pp. 12\,382--12\,395, 2020.

\bibitem{HJ:01:GMD}
H.~Jaeger, ``The {“echo state”} approach to analysing and training
  recurrent neural networks-with an erratum note,'' \emph{” GMD - German
  National Research Institute for Computer Science, Tech. Rep.}, vol. 148,
  no.~34, p.~13, 2010.

\bibitem{Lillicrap:15}
T.~P. Lillicrap \emph{et~al.}, ``Continuous control with deep reinforcement
  learning,'' in \emph{Proc. 4th International Conf. on Learning
  Representations (ICLR)}, 2016.

\bibitem{Wang:15}
\BIBentryALTinterwordspacing
Z.~Wang, T.~Schaul, M.~Hessel, H.~van Hasselt, M.~Lanctot, and N.~de~Freitas,
  ``Dueling network architectures for deep reinforcement learning,'' 2015.
  [Online]. Available: \url{arXiv preprint arXiv:1511.06581}
\BIBentrySTDinterwordspacing

\bibitem{Puterman:14}
M.~L. Puterman, \emph{{Markov} Decision Processes: Discrete Stochastic Dynamic
  Programming}.\hskip 1em plus 0.5em minus 0.4em\relax John Wiley \& Sons,
  Inc., 1994.

\bibitem{BD:95:Book:v1}
D.~P. Bertsekas, \emph{Dynamic programming and optimal control}.\hskip 1em plus
  0.5em minus 0.4em\relax Athena scientific Belmont, MA, 1995, vol.~1, no.~2.

\bibitem{Abadi:16}
M.~Abadi \emph{et~al.}, ``{Tensorflow}: A system for large-scale machine
  learning,'' in \emph{Proc. 12th USENIX Sym. Opr. Syst. Design and Imp. (OSDI
  16)}, Nov. 2016, pp. 265--283.

\bibitem{DJ:14}
\BIBentryALTinterwordspacing
D.~P. Kingma and J.~L. Ba, ``{Adam}: A method for stochastic optimization,''
  2014. [Online]. Available: \url{arXiv preprint arXiv:1412.6980}
\BIBentrySTDinterwordspacing

\end{thebibliography}
\end{document}